\documentclass[%
 aip,
 amsmath,amssymb,
preprint,%
]{revtex4-1}

\usepackage{graphicx}% Include figure files
\usepackage{dcolumn}% Align table columns on decimal point
\usepackage{bm}% bold math
\usepackage[utf8]{inputenc}
\usepackage[T1]{fontenc}
\usepackage{mathptmx}
\usepackage{textcomp}
\usepackage{dsfont}
\usepackage{subfig}
\usepackage{multirow}
\usepackage{xcolor}

\begin{document}

%\preprint{AIP/123-QED}

\title[Vorticity Dynamics on a Sphere with DEC]{Effects of Rotation on Vorticity Dynamics on a Sphere with Discrete Exterior
Calculus}
% Force line breaks with \\

\author{Pankaj Jagad}
\email{pankaj.jagad@kaust.edu.sa}
 \affiliation{Mechanical Engineering, Physical Science and
%\affiliation{Physical Science and
Engineering Division, King Abdullah University of Science and Technology,
Thuwal, Saudi Arabia}%Lines break automatically or can be forced with \\

\author{Ravi Samtaney}
 %\homepage{https://urldefense.com/v3/__http://www.Second.institution.edu/*Charlie.Author__;fg!!Nmw4Hv0!kenZHjs85jWuRmfG3Ru4Lz6g-7DbbZURy9kr-dJGXc0-7lwPJ1OXJK1n1Is3w8YjkcN_Zog$ .}
\email{ravi.samtaney@kaust.edu.sa}
\affiliation{%
Mechanical Engineering, Physical Science and
%Physical Science and
Engineering Division, King Abdullah University of Science and Technology,
Thuwal, Saudi Arabia%\\This line break forced% with \\
}%

\date{\today}% It is always \today, today,

\begin{abstract}
 
We investigate incompressible, inviscid vorticity dynamics on a rotating unit sphere using a Discrete Exterior
Calculus (DEC) scheme. 
For a prescribed initial vorticity distribution,
we vary the rate of rotation of the sphere from zero (non-rotating
case, which corresponds to infinite Rossby number (Ro)) to 320 (which
corresponds to Ro = $1.30 \times 10^{-3}$), and investigate the evolution with time
of the vorticity field. For the non-rotating case, the vortices evolve
into thin filaments due to so-called forward/direct enstrophy cascade.
 At late times an oscillating quadrupolar vortical
field emerges as a result of the inverse energy cascade. 
Rotation diminishes the forward cascade of enstrophy (and hence the inverse cascade of energy)
and tend to align the vortical structures in the azimuthal/zonal direction.
Our investigation reveals that, for the initial vorticity field comprising of intermediate-wavenumber spherical harmonics, the zonalization of
the vortical structures is not monotonic with ever decreasing
Rossby numbers and the structures revert back to a non-zonal state
below a certain Rossby number. On the other hand, for the initial vorticity field comprising of intermediate to large-wavenumber spherical harmonics, the zonalization is monotonic with decreasing Rossby number. Although, rotation diminishes the forward
cascade of enstrophy, it does not completely cease/arrest the cascade. 
\end{abstract}

\maketitle

%\begin{quotation}
%The ``lead paragraph'' is encapsulated with the \LaTeX\ 
%\verb+quotation+ environment and is formatted as a single paragraph before the first section heading. 
%(The \verb+quotation+ environment reverts to its usual meaning after the first sectioning command.) 
%Note that numbered references are allowed in the lead paragraph.
%
%The lead paragraph will only be found in an article being prepared for the journal \textit{Chaos}.
%\end{quotation}

\section{\label{sec:intro}Introduction\protect\\}

It is a common practice to investigate planetary fluid dynamics assuming the flow to be incompressible and inviscid. 
Such flows are typically characterized by two-dimensional
(2D) turbulence wherein there is direct cascade of enstrophy to smaller
scales and inverse cascade of energy to larger scales. Based on a
2D turbulence model, Lorenz\cite{doi:10.3402/tellusa.v21i3.10086}
predicted the atmospheric flow as a function of spatial scales. Leith\cite{1971JAtS...28..145L}, and  Leith \& Kraichnan\cite{Leith1972}
refined Lorenz\textquoteright s predictability estimates to further
illustrate the application of a turbulence theory to important meteorological
problems. In fact, the topic of 2D turbulence dates back to the theoretical
studies of Kraichnan\cite{doi:10.1063/1.1762301}, Kraichnan \& Montgomery\cite{Kraichnan_1980},
and Leith\cite{doi:10.1063/1.1691968} on an infinite plane. They
predicted the classical power laws of $k^{-3}$ in the forward enstrophy-cascade
range and of $k^{-5/3}$ in the inverse energy-cascade range for
the energy spectrum. Lilly\cite{doi:10.1063/1.1692444} numerically
integrated the 2D incompressible Navier-Stokes equations in order
to test the validity of Kraichnan's predictions on the structure of
2D turbulence. Batchelor\cite{doi:10.1063/1.1692443} computed the
energy spectrum in homogeneous 2D turbulence. Newell\cite{newell_1969} proposed a mechanism comprising of the
resonant interaction of Rossby wave packets to generate planetary
zonal flows. However, these studies do not examine the reasons for the existence of Rossby waves. 
Furthermore, these classical studies
did not consider the effect of planetary rotation ("beta"), and
the presence of long-lived coherent vortices in 2D turbulence was
not widely known \citep{Huang1998}.

\subsection{2D turbulence on a $\beta$-plane}
Rhines\cite{rhines_1975} presented the first
geophysical application of the 2D turbulence theory. The effect of
planetary rotation on 2D turbulence was investigated with
a numerical model on a $\beta$-plane. The inverse energy cascade
was shown to cease roughly at a characteristic wave number $k_{\beta}=\sqrt{\beta/2U}$
(also known as the Rhines scale $k_{R}$), where $k$ is the total wave
number, $U$ is the RMS velocity and $\beta$ is the meridional gradient
of the Coriolis parameter (later denoted by $f$). The turbulence transforms into Rossby
waves around the wave number $k_{\beta}$. The $\beta$-effect (the
meridional variation of the Coriolis parameter $\beta$) causes the flow field to be anisotropic,  and a zonal band structure consisting
of alternating easterly and westerly jets emerges. The stronger
$\beta$-effect on eddies that are meridionally elongated compared with zonally elongated
eddies makes the Rhines scale anisotropic. 
Holloway \& Hendershott\cite{holloway_hendershott_1977} investigated decaying 2D turbulence
on a $\beta$-plane and observed zonal anisotropy for wavenumber
$k\leq k_{\beta}^{H}\equiv\beta/Z$, where $Z$ is the RMS vorticity.
Shepherd\cite{shepherd_1987} extended the theory of homogeneous
barotropic $\beta$-plane turbulence to include effects arising from spatial
inhomogeneity in the form of a zonal shear flow and demonstrated profound
effects of the background shear flow on the inhomogeneous turbulence.
The shear straining process induced a downscale enstrophy transfer
similar to the traditional downscale enstrophy cascade due to local
eddy-eddy interaction in spectral space. The shear was shown to induce
transfer of disturbance energy into the range of $k\leq k_{\beta}$
and make the disturbance flow field to become meridionally anisotropic
in the low-wave number range. The observed atmospheric energy spectrum
was explained with these concepts \citep{1987JAtS...44.1166S}. Yamada \& Yoneda\cite{YAMADA20131} proved, with a mathematically
rigorous theorem, that at a high $\beta$, the resonant interactions
of Rossby waves (which are expected to dominate the $\beta$-plane
dynamics) govern the flow dynamics for an incompressible 2D flow on
a $\beta$-plane.
Several studies\cite{10.1175/1520-0485,maltrud_vallis_1991,doi:10.1063/1.1327594,doi:10.1063/1.1327594,CHEKHLOV1996321,Basdevant1981,holloway_hendershott_1977,doi:10.1080/03091928008241180,bartello_holloway_1991,10.1175/1520-0485,doi:10.2514/5.9781600866340.0108.0120,Panetta1993,doi:10.1063/1.166007,doi:10.1063/1.166011,10.1175/JAS4013.1} present the dynamics of forced 2D turbulence (such as ``wave-turbulence boundary", zonostrophic turbulence, etc.)  on a $\beta$-plane.

\subsection{2D turbulence on a rotating sphere\label{sec:decaying_2d_turbulence}}
Yoden \& Yamada\cite{Yoden1993}
investigated the effects of rotation and sphericity on decaying
2D turbulence on a rotating sphere. Large rotation rates, under the
existence of Rossby waves, revealed an easterly jet at high latitudes.
Huang \& Robinson\cite{Huang1998}
examined the dynamics of 2D turbulence on a rotating sphere, and 
derived and verified the anisotropic Rhines scale in decaying turbulence
simulations. The inverse energy cascade along the zonal axis (zonal
wavenumber $m=0$) was not directly arrested by beta in their simulations. 
Coherent polar vortices emerge in decaying 2D turbulence simulations
on a rotating sphere, although multiple zonal jets are difficult to
obtain \citep{Yoden1993}. However,
rotating shallow-water turbulence with a finite radius of deformation
does generate multiple zonal jets \citep{doi:10.1063/1.868929}. For
a variety of initial conditions and a sufficiently large rotation rate,
a band structure at mid-latitudes and westward circumpolar jets
in the polar regions have been observed \citep{YODEN1999,Hayashi2000,ishioka:hal-00301974}.
Hayashi {\it et al.}\cite{10.1175/2007JAS2209.1} reviewed jet formation
/ zonal mean flow generation in decaying 2D turbulence
on a rotating sphere from the view-point of Rossby waves with 
two parameter space of the rotation rate and Froude number Fr. For
a nondivergent flow (low Fr) and large rotation rate, intense easterly
circumpolar jets in addition to a banded structure of zonal mean flows
with alternating flow directions were found. However, for divergent
flows (with increasing Fr), circumpolar jets disappear and an equatorial
easterly jet emerges. Takehiro {\it et al.}\cite{Takehiro_2007a} investigated
the strength and width of the easterly circumpolar jets and discovered
asymptotic behaviors in rapidly rotating cases. Extremely inhomogeneous
banded structure of zonal flows and accumulation of most of the kinetic
energy inside the easterly circumpolar jets were revealed. Takehiro {\it et al.}\cite{Takehiro_2007b} confirmed these revelations
by performing 2D decaying turbulence simulations for a barotropic fluid on a rotating
sphere. Establishment of the banded structure of zonal flows was observed
relatively early in the simulations.   
At late times, only the
circumpolar jets were intensified gradually, while no further evolution
in the banded structure in the low and midlatitudes was observed.
The easterly momentum transport from the low and midlatitudes associated
with Rossby waves contributes to the maintenance of the circumpolar
easterly jets \citep{Hayashi2000,10.1175/2007JAS2209.1}. Yoden {\it et al.}\cite{10.1007/978-4-431-67002-5_22} investigated 2D
decaying turbulence for a nondivergent barotropic fluid on a rotating
sphere to survey the nature of pattern formation from random initial
fields. Isolated coherent vortices emerged in non-rotational cases
as in the planar 2D turbulence. However, a westward circumpolar vortex
in high-latitudes and zonal band structures in mid and low-latitudes
appeared with increasing rotation rate. They investigated dependence
of these features on the initial energy spectrum and discussed the
dynamics of such pattern formations with a weakly nonlinear Rossby
wave-zonal flow interaction theory. Sasaki {\it et al.}\cite{sasaki_takehiro_yamada_2012}
examined the stability of inviscid zonal jet flows on a rotating sphere. Saito \& Ishioka\cite{Izumi_SAITO20162016-002}
obtained a quasi-invariant associated with the emergence of
zonally elongated structures in 2D turbulence on a rotating sphere
by a minimization process. They explained the anisotropic energy transfer
that favors zonally elongated structures depicting airfoil-shaped contours
for the weighting coefficient distribution. The quasi-invariant (defined as a weighted sum of the energy
density in the wavenumber space) 
was shown to conserve well if nonlinearity
of the system is sufficiently weak. Obuse \& Yamada\cite{PhysRevFluids.4.024601} investigated
three-wave resonant interactions of Rossby-Haurwitz waves in 2D turbulence
on a rotating sphere. According to them, the zonal waves of the form
$Y_{l}^{m=0}\exp\left(i\omega t\right)$ with odd $l$ should be considered
for inclusion in the resonant wave set to ensure that the dynamics
of the resonant wave set determine the overall dynamics of the turbulence
on a rapidly rotating sphere. Here, $Y_{l}^{m}$ are the spherical
harmonics and $\omega=-2\Omega m/\left[l\left(l+1\right)\right]$
is the frequency of a Rossby-Haurwitz wave.

\subsubsection{Shallow-water turbulence}
The shallow-water equations are the simplest model of planetary flows
considering the effects of divergence \citep{farge_sadourny_1989,yuan_hamilton_1994}.
Cho \& Polvani\cite{doi:10.1063/1.868929} found remarkably different
characteristic flow patterns in shallow-water turbulence from that
in 2D non-divergent turbulence. Their investigations revealed a retrograde
equatorial jet instead of a polar jet (which is dominant in 2D turbulence)
because of the effects of planetary rotation. The asymmetry between
a cyclone and an anticyclone is attributed to the predominance of a retrograde
jet in the shallow-water system \citep{doi:10.1063/1.869898}. Kitamura \& Ishioka\cite{10.1175/JAS4015.1} performed ensemble experiments
of decaying shallow-water turbulence on a rotating sphere to confirm
the robustness of the emergence of an equatorial jet. Predominance
of a prograde jet, although less likely in shallow-water turbulence,
was also noted. From the examination of a zonal-mean flow induced by wave-wave
interactions, using a weak nonlinear model, they found that Rossby
and mixed Rossby-gravity waves induce second-order acceleration.
 Yoden {\it et al.}\cite{10.1007/978-94-007-0360-5_21}
reviewed jet formation in decaying 2D turbulence on a rotating sphere
considering wave mean-flow interaction for both shallow-water case
and non-divergent case (in the limit Fr $\rightarrow$ zero). They investigated
the behavior of mean zonal flow generation in the two dimensional ($\Omega$, Fr) parameter space
where $\Omega$ is the non-dimensional rotation rate. For the non-divergent
flow and large $\Omega$, an intense retrograde circumpolar jet and a
banded structure of mean zonal flows with alternating flow directions
in middle and low latitudes emerged. With increasing Fr, the circumpolar
jets disappeared and a retrograde jet emerged in the equatorial region.
The appearance of the intense retrograde jets was attributed to the
angular momentum transport associated with the propagation and absorption
of Rossby waves. In non-divergent flows, long Rossby waves tend to
be absorbed around the poles. However, for large Fr, Rossby waves
hardly propagate towards the poles and are absorbed near the equator.
The equatorial jet was not always retrograde, the emergence of a less
likely prograde jet was also found. 

Although investigating the dynamics of {\em forced} 2D turbulence on a rotating sphere is beyond the scope of the present work, we, nonetheless, note several interesting studies\cite{Williams1978,Basdevant1981,doi:10.1063/1.869518,doi:10.1063/1.869327,Huang1998,doi:10.1063/1.34273,doi:10.1063/1.1327594,PhysRevLett.89.124501,10.1175/JAS4003.1,doi:10.1063/1.3407652,10.1175/1520-0485,doi:10.1063/1.1373684,doi:10.1029/2004GL019691,galperin:hal-00302701,Galperin_2008,vallis_2017,CHEKHLOV1996321,doi:10.1029/2004GL020106,galperin:hal-00302701,10.1175/2007JAS2219.1,PhysRevLett.101.178501}.

\subsection{Discrete exterior calculus (DEC)}

Before delving into some conclusions drawn from the review, we digress to present a brief introduction of DEC. Exterior calculus deals with the calculus of differential geometry,
hence with the calculus of differential forms, and provides an alternative
to the vector calculus. DEC is a numerical exterior calculus and deals
with the discrete differential forms. A discreet differential form
is an integral quantity on a mesh object, e.g., integral of a vector
along a mesh edge $\int\mathbf{v}\cdot\mathbf{dl}$ represents a discrete
1-form. DEC retains at the discrete level many of the identities of
its continuous counterpart. It is coordinate independent, therefore
suitable for solving flows over curved surfaces. For the further reading, a few representative DEC references include the references \onlinecite{flanders1963differential, abraham2012manifolds, hirani2003discrete, desbrun2003discrete, desbrun2005discrete, grinspun2006discrete, hirani2008numerical, desbrun2008discrete, crane2013digital, perot2014differential, mohamed2016discrete, de2016subdivision}.

\subsection{Scope of the present study}
The literature suggests that there is no uniform agreement about the arrest of the cascade at the Rhines scale. Moreover, the focus of the previous investigations was to analyze the effect of rotation on the inverse energy cascade, and the effect on the direct enstrophy cascade was implied. A comprehensive analysis investigating the effect of rotation on the vorticity dynamics on a sphere is warranted.  We investigate the effect of rotation on vorticity dynamics on a unit
sphere (which falls in the category of decaying 2D turbulence as reviewed in section \ref{sec:decaying_2d_turbulence}) using a DEC method \citep{jagad2020primitive}. The evolution of the 
vorticity field from an arbitrary initial vorticity distribution,
is examined with an emphasis on investigating the effect of rotation.  Presently, we vary the rate of rotation
of the sphere $\left(\Omega\right)$ from zero (Ro = $\infty$) to 320 (Ro = $1.30 \times 10^{-3}$). Here, the Rossby number $\mathrm{Ro}=U/2\Omega L $, where $U$ is the characteristic velocity scale (which is assumed to be equal to square root of total kinetic energy), $L$ is the characteristic length scale (assumed as the radius of sphere). The sphere undergoes $\Omega/\left(2\pi\right)$ rotations per unit time. Considering $U\approx 10$ m/s for the atmosphere,  and $L\approx 6 \times 10^{6}$ m, then $\Omega = 36.47$ rad/s for the unit sphere represents the rotating earth. Moreover, we choose the eddy turnover time ( = $4\pi / \left | \omega_{max} \right|$ with $\left | \omega_{max} \right |$ denoting the maximum vorticity magnitude) as the characteristic time scale for the representation of non-dimensional time $t$. In addition, we vary the range of spherical harmonics wave numbers (with total initial kinetic energy held approximately constant) constituting the initial vorticity field, and investigate the effect of differential initial spectral representations. Table \ref{tab:Simulation-parameters} shows the parameters employed in the present study. We vary the initial condition from  case A to F, and for each case, we vary Ro as indicated in table \ref{tab:Simulation-parameters}. 

As discussed in detail later, the
time evolution of the vorticity field at different Ro indicates that
increasing rotation diminishes the forward cascade of enstrophy and zonalizes
the vortical structures. However, the zonalization of the structures
does not continue monotonically with ever decreasing Rossby numbers,
and the structures tend to a non-zonal state below a certain
Rossby number for the initial vorticity field comprising
of intermediate-wavenumber spherical harmonics (for test cases A, B, and C). Whereas, for the initial vorticity field comprising also of large-wavenumber spherical harmonics (for test cases D, E, and F), the tendency to zonalization is monotonic. Furthermore, we express the vorticity distribution in terms of spherical harmonics, and determine the spherical harmonic modes. From the spectral content, we determine the vorticity power spectrum and the effect of rotation on it. Moreover, we investigate the effect of rotation on the spectral distribution of the vorticity power. We also determine the effect of rotation on the vorticity probability density, which is a measure of the area occupied by each vorticity level. Additionally, we identify individual vortical structures using a segmentation procedure, and determine the effect of rotation on the probability density of aspect
ratio (ratio of major to minor axis length of the vortical structures),
and that of the cosine of the angle between the major axis and the
azimuthal direction (which is a measure of orientation of the vortical
structures). The analysis further confirms the non-monotonic and monotonic nature of zonalization
with decreasing Ro observed qualitatively in the time evolution of
the vorticity field, respectively for intermediate-wavenumbers and intermediate to large-wavenumbers constituting the initial vorticity field. Moreover, it reveals that although, rotation
diminishes the forward cascade of enstrophy, it does not completely
cease/arrest the cascade. 

The outline of the paper 
is as follows. The simulations results are
presented along with detailed quantification of the important
dynamical quantities first. The paper closes by emphasizing the key
conclusions deduced from the conducted simulations and analysis of the results. A brief description of the physical setup and numerical procedure is presented in appendix \ref{sec:num_proc}, followed by a case for the verification of the numerical procedure in appendix \ref{appB0}. The segmentation algorithm employed for identifying individual vortical
structures is presented in appendix \ref{appB}.

%\begin{center}
\begin{table*}

\centering{}%
\begin{tabular}{|ccccccc|}
\hline 
&\multicolumn{6}{c|}{Test case} \tabularnewline
&\multirow{1}{*}{A} & \multirow{1}{*}{B} & \multirow{1}{*}{C} & \multirow{1}{*}{D} & \multirow{1}{*}{E} & \multirow{1}{*}{F}\tabularnewline
\hline 
Spherical harmonics wavenumber $\left(l\right)$ range & \multirow{2}{*}{4 - 10} & \multirow{2}{*}{4 - 10 } & \multirow{2}{*}{4 - 10} & \multirow{2}{*}{4 - 20} & \multirow{2}{*}{4 - 40} & \multirow{2}{*}{4 - 80}\tabularnewline
%Spherical harmonics wavenumber $\left(l\right)$ range & {4 - 10} & {4 - 10 } & {4 - 10} & {4 - 20} & {4 - 40} & {4 - 80}\tabularnewline
constituting the initial vorticity field  &  &  &  &  &  & \tabularnewline
\hline 
\multirow{8}{*}{Rossby number (Ro)} & \multicolumn{6}{c|}{$\infty$}\tabularnewline
%\cline{2-7} 
 & \multicolumn{6}{c|}{$8.34\times10^{-2}$}\tabularnewline
%\cline{2-7} 
 & \multicolumn{6}{c|}{$4.17\times10^{-2}$}\tabularnewline
%\cline{2-7} 
 & \multicolumn{6}{c|}{$2.08\times10^{-2}$}\tabularnewline
%\cline{2-7} 
 & \multicolumn{6}{c|}{$1.04\times10^{-2}$}\tabularnewline
%\cline{2-7} 
 & \multicolumn{6}{c|}{$5.20\times10^{-3}$}\tabularnewline
%\cline{2-7} 
 & \multicolumn{6}{c|}{$2.60\times10^{-3}$}\tabularnewline
%\cline{2-7} 
 & \multicolumn{6}{c|}{$1.30\times10^{-3}$}\tabularnewline
\hline 
\end{tabular}
\begin{centering}
\caption{Simulation parameters. \label{tab:Simulation-parameters}. Note that spherical harmonics wavenumber $\left(l\right)$ range constituting the initial vorticity field is the same for the cases A, B, and C, but the amplitude is different.} 
\par\end{centering}
\end{table*}
%\par\end{center}

\section{Results and Discussion}

\begin{figure}
\centering{}\includegraphics[scale=0.18]{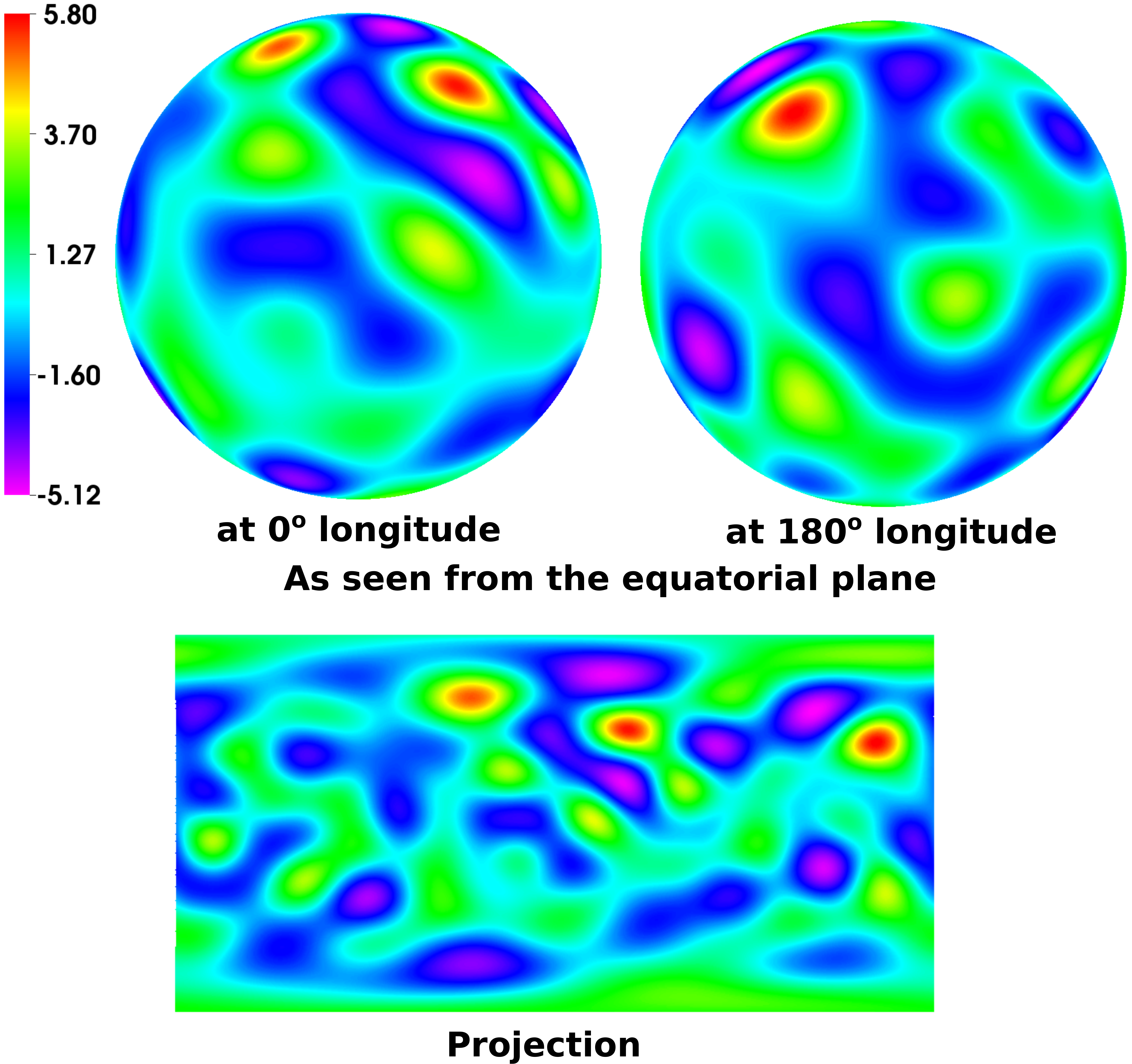}\caption{Arbitrary initial vorticity field computed from the spherical harmonic modes. The bottom panel shows the spherical surface projected using the standard orthographic projection. 
\label{fig:Initial-conditions}}
\end{figure}

In this section, we first present the results for one of the cases (case A) with arbitrary initial vorticity field comprising of spherical harmonics wavenumbers $l = 4 - 10$ (see table \ref{tab:Simulation-parameters}) . This is followed by a discussion of other cases emphasizing the effect of wavenumber range comprising the initial vorticity field. 
We compute the initial stream function distribution at the primal mesh nodes as
 
\begin{equation} 
\psi\left(\theta,\phi\right)=\sum_{l=4}^{10}\sum_{m=-l}^{l}\psi_{lm}Y_{l}^{m}\left(\theta,\phi\right), 
\label{eq:shexpansion} 
\end{equation} 
\noindent where $\theta$ is the colatitude, $\phi$ is the longitude, $Y_{l}^{m}$ is the spherical harmonic function of degree $l$ and order $m$, and $\psi_{lm}$ are the expansion coefficients (or spherical harmonic modes) with $\psi_{l-m}=\left(-1\right)^{m}\psi_{lm}^{\ast}$. Here, $\psi_{lm}^{\ast}$ is the complex conjugate of $\psi_{lm}$. The coefficients $\psi_{lm}$ are assigned values as given in Dritschel {\it et al.}\cite{dritschel2015late}. The degree $l$ characterizes the total wavenumber and order $m$ characterizes the azimuthal/zonal wavenumber of the spherical harmonics. The initial mass flux 1-form for a primal mesh edge is now computed as $u^{\ast}=d_{0}\psi$. The vorticity distribution (that of the component of vorticity vector normal to the surface) at the primal mesh nodes is computed as $\omega=\ast_{0}^{-1}\left[-d_{0}^{T}\right]\ast_{1}u^{\ast}$. Figure \ref{fig:Initial-conditions} shows the initial vorticity distribution. 

\subsection{Vorticity field evolution}
\begin{figure*}

\centering{}\includegraphics[scale=0.2]{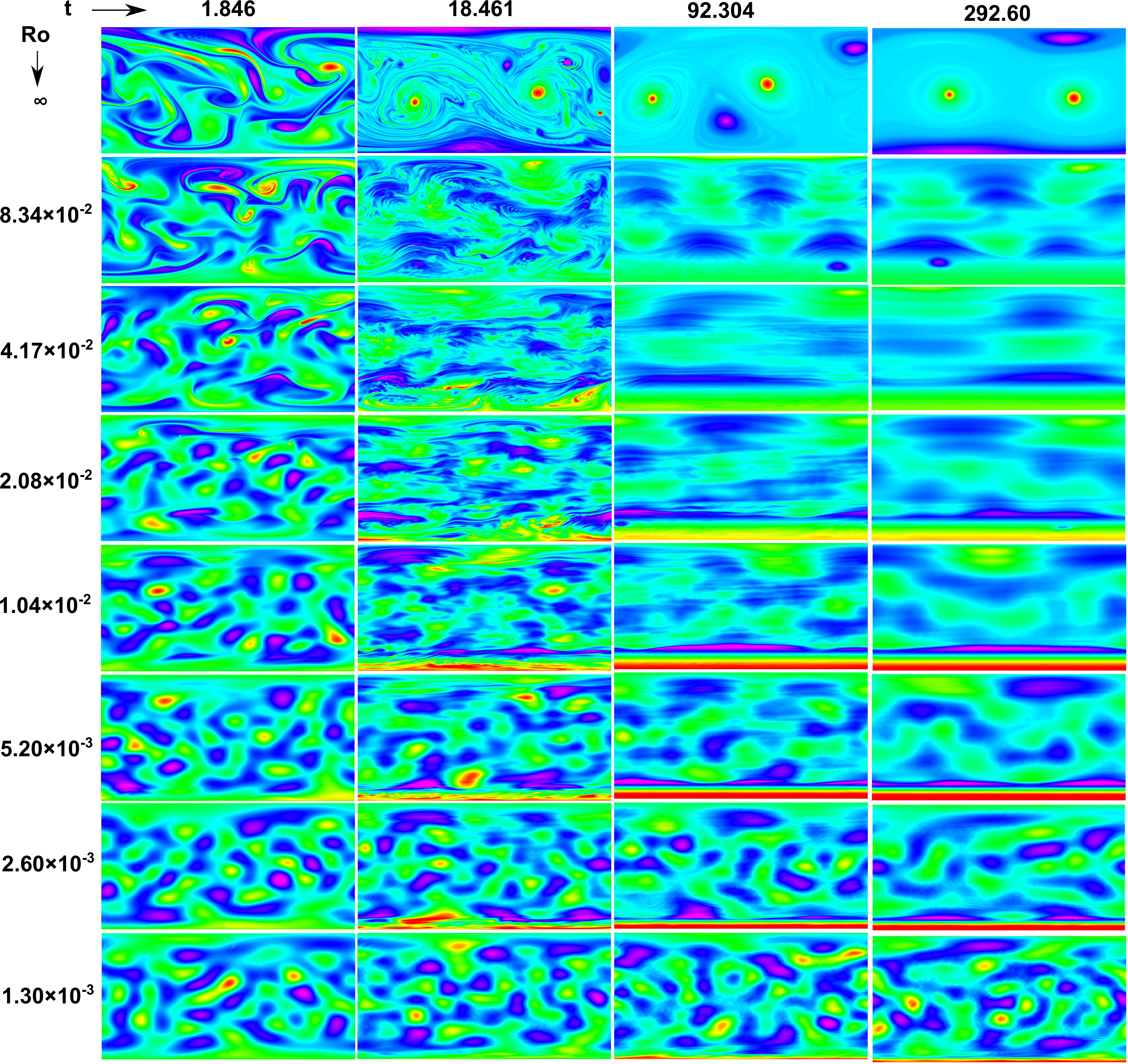}\caption{Time evolution of vorticity field as a function of Ro, showing diminishing enstrophy cascade and zonalization of vortical structures with decreasing Ro.\label{fig:Vorticity-distribution-at}}
\end{figure*}

Figure \ref{fig:Vorticity-distribution-at} shows the vorticity distribution with varying  Rossby numbers at four time instants (early, mid, and late-to-very late).  For the non-rotating case $\mathrm{Ro}=\infty$ (top panel in Figure \ref{fig:Vorticity-distribution-at}) the vortices rapidly develop thin filaments due to the forward enstrophy cascade by time $t=18.461$  (eddy turnover time). 
Smaller scale vortices merge to form larger ones due to the inverse energy cascade.
At late times ($t=92.304$, and $t=292.60$) much of the enstrophy has cascaded beyond the smallest scale resolved in the simulation. Due to the biharmonic term, the grid level small scale enstrophy dissipates numerically. An oscillating quadrupolar vortical structure, similar to that in Dritschel {\it et al.}\cite{dritschel2015late}, is the late time outcome. 

On the other hand, for the rotating cases (finite $\mathrm{Ro}$ number) we note that the evolution into the thin filaments in inhibited and the absence of the quadrupolar vortical structure at late times. Thus, rotation diminishes the forward cascade of enstrophy (and also the inverse cascade of energy). Moreover, as the Rossby number decreases, up to Ro $=2.08 \times 10^{-2}$, the vortices tend to become elongated and aligned along the azimuthal direction (tend to become zonal - a process dubbed as ``zonalization"). With increased rotation, i.e., further decrease in Ro, these structures tend towards smaller elongation and non-zonal in character. Thus, the zonalization of the structures with decreasing Ro is non-monotonic (for the present case) and this is apparent at  late times. We also note the presence of circumpolar vortices for the rotating cases as previously reported \citep{Yoden1993,doi:10.1063/1.868929}. 

In order to quantify the effect of rotation, vorticity power spectra, spectral distribution of vorticity power, vorticity probability density, probability density  of aspect ratio (ratio of major to minor axis length of vortical structures) and the cosine of the angle between the major axis and the azimuthal direction  (a measure of orientation of the vortical structures) are determined. These quantifications are discussed subsequently.
\subsection{Vorticity power spectra}
Spherical harmonics  satisfying the following orthogonality relation \citep{Wieczorek2018}
\begin{equation} 
\int_{\Omega}Y_{lm}\left(\theta,\phi\right)Y_{l'm'}\left(\theta,\phi\right)\mathrm{d}\Omega=4\pi\delta_{ll'}\delta_{mm'}, 
\end{equation} 
\noindent where $\Omega$ is the surface area of the sphere, and $\delta_{ij}$ is the Kronecker delta function. With this, the spherical harmonic expansion coefficients are expressed as
\begin{equation} 
\omega_{lm}=\frac{1}{4\pi}\int_{\Omega}\omega\left(\theta,\phi\right)Y_{lm}\left(\theta,\phi\right)\mathrm{d}\Omega. 
\end{equation}
The total vorticity power in the physical and spectral domains is related as
\begin{equation} 
\frac{1}{4\pi}\int_{\Omega}\omega^{2}\left(\theta,\phi\right)\mathrm{d}\Omega=\sum_{l=0}^{\infty}\sum_{m=-l}^{l}\omega_{lm}^{2}=\sum_{l=0}^{\infty}S_{ff}\left(l\right), \label{eq:sffl} \end{equation} 
\noindent where  $S_{ff}\left(l\right)$ is the power spectrum of vorticity. We use Eq. (\ref{eq:sffl}) to compute the vorticity power spectrum $S_{ff}\left(l\right)$.

\begin{figure*}
\centering{}
\subfloat[]{\begin{centering}
\includegraphics[scale=0.35]{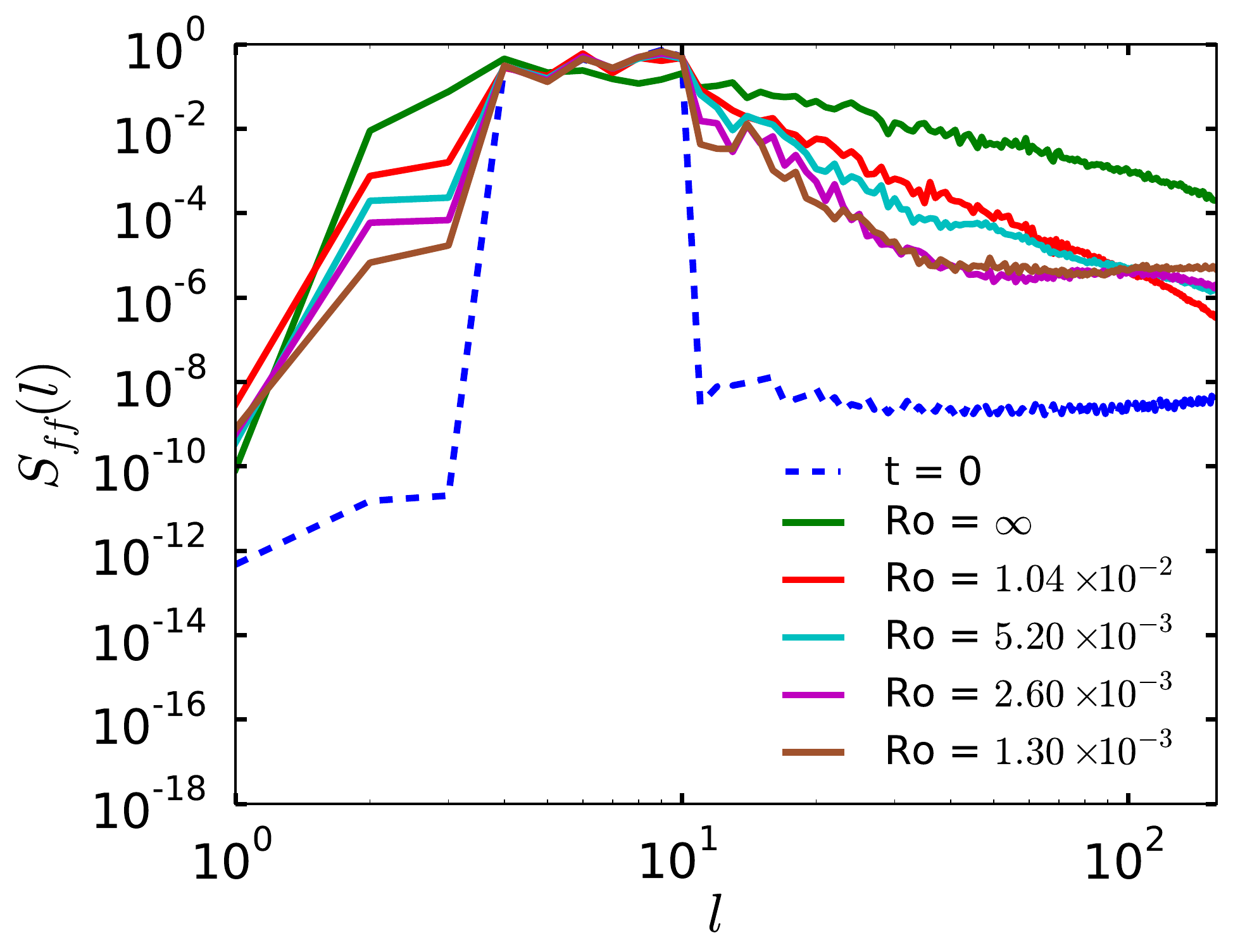}
\par\end{centering}
}\subfloat[]{
\centering{}\includegraphics[scale=0.35]{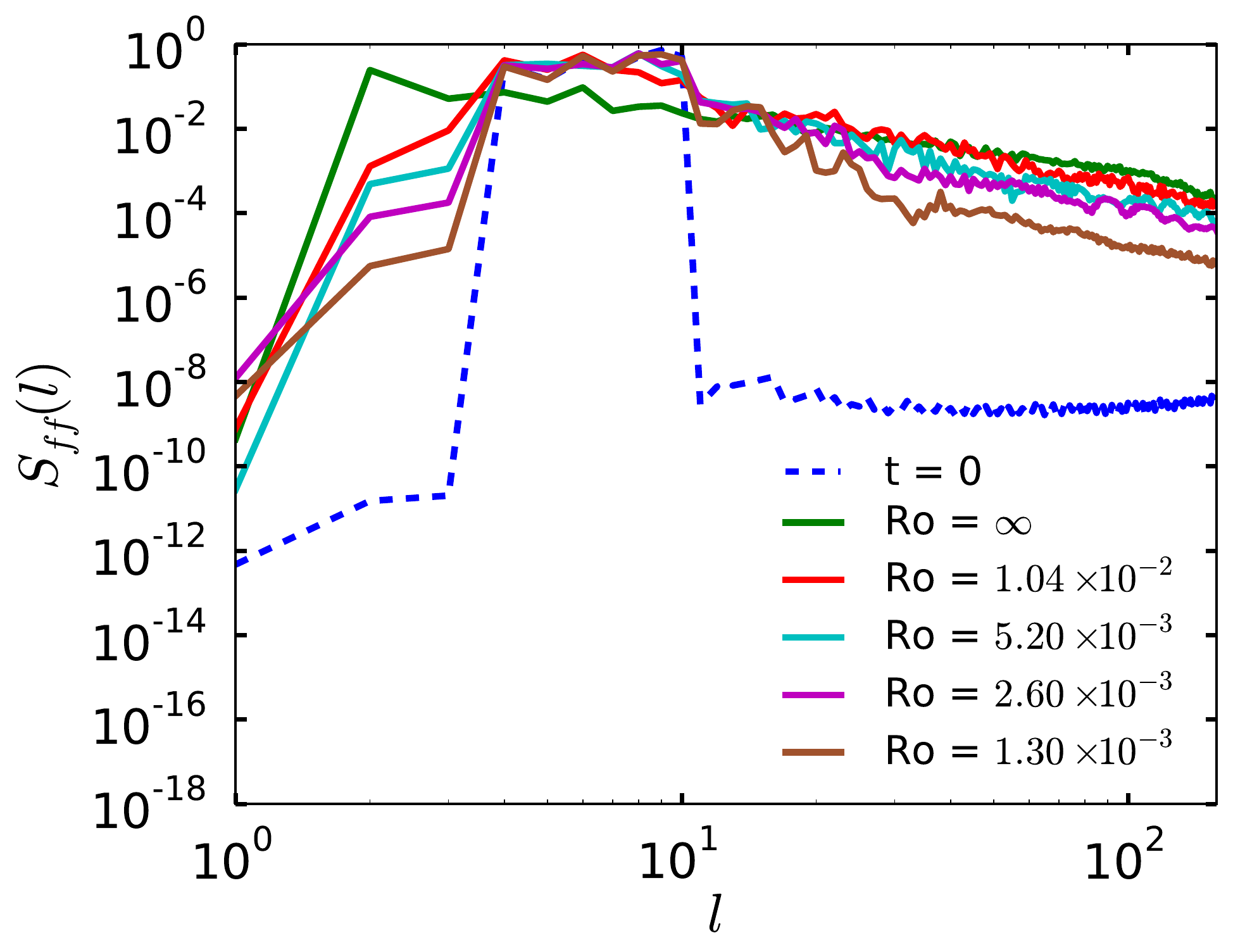}} \\
\subfloat[]{
\centering{} \includegraphics[scale=0.35]{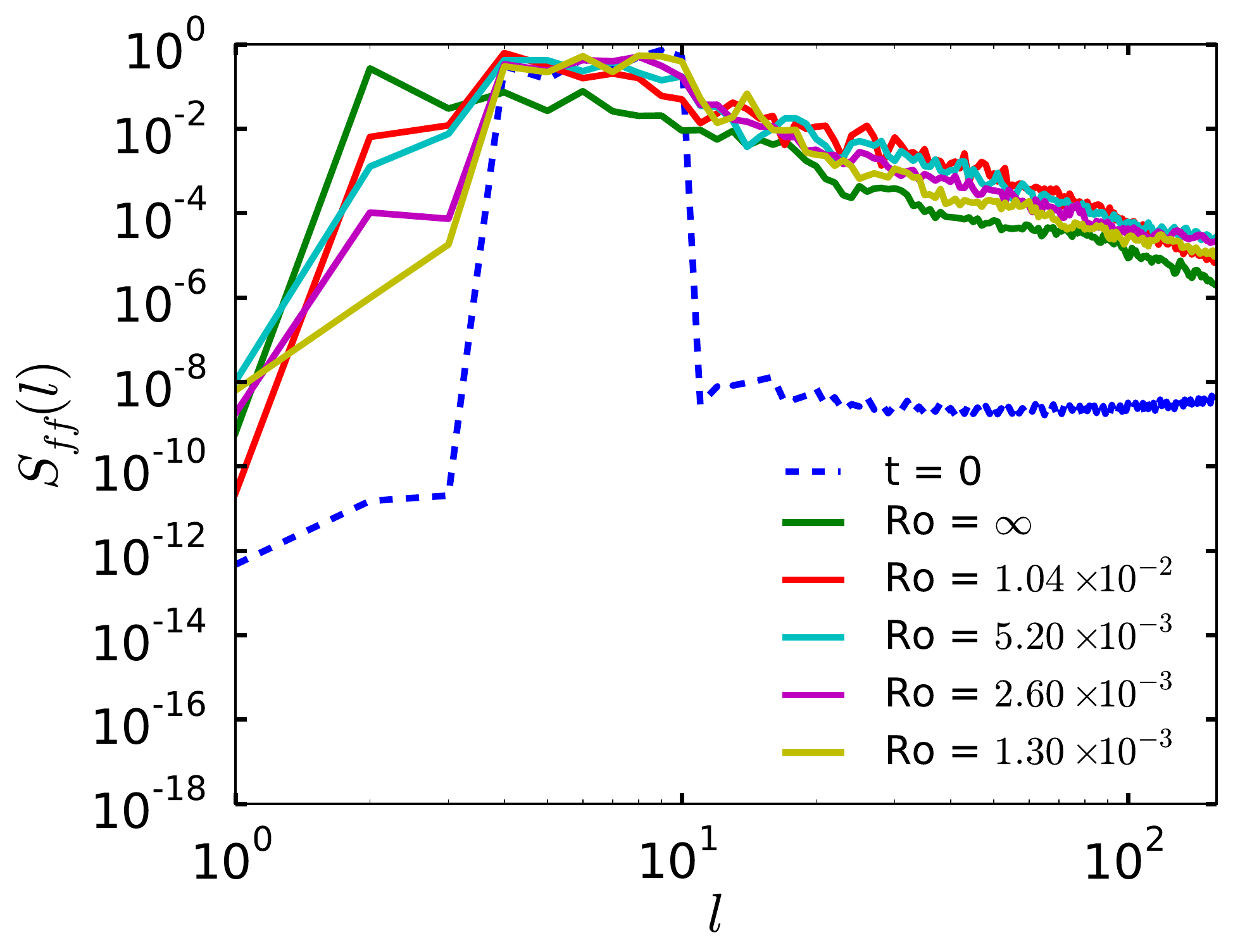}}
\caption{Vorticity power spectra  as a function of wavenumber for varying Ro, depicting diminishing enstrophy cascade with decreasing Ro. Time instants shown are  (a) $t=1.846$, (b) $t=18.461$, (c) $t=92.304$. \label{fig:Vorticity-power-spectra}}
\end{figure*}

Figure \ref{fig:Vorticity-power-spectra} shows vorticity power spectra for three time instants and for different rotation rates. The initial (at t = 0) vorticity power is confined to a wavenumber range of $l=4-10$. This power cascades to the larger waver numbers as the flow evolves with time. As the Rossby number decreases (increase in rotation rate), the slope of the vorticity power spectra increases. This implies that the forward cascade of enstrophy diminishes with decreasing Rossby number. With decreasing Ro, the proportion of the spectral power contained in the wave numbers $l=4-10$ increases, even at a late time t = 92.304, further supporting the observation that the enstrophy cascade diminishes with decreasing Ro. However, there is still significant power contained in the larger wave numbers, even at a very small Ro = $1.30 \times 10^{-3}$, showing that the forward enstrophy cascade is not  suppressed/arrested completely which is somewhat contrary to the previously held notions of the enstrophy cascade \citep{rhines_1975,Yoden1993,doi:10.1063/1.1327594}.   

\subsection{Spectral distribution of vorticity power}
It is interesting to examine the effect of rotation on the distribution of the vorticity power in spectral space (i.e. using the  $m-l$ space of spherical harmonics). The results are plotted in figure \ref{fig:Vorticity-power-distribution} where the spectral distribution is shown using right angle triangles: the vertical (resp. horizontal) edge of each triangle is the total wavenumber $l$ (azimuthal wave number $m$, resp.).  
In figure \ref{fig:Vorticity-power-distribution}(a) we plot the 
initial vorticity power, comprising 
of the total modes $l=4-10$ and all of the initial zonal modes $m=4-10$,
and is isotropic
with respect to the zonal modes. 
The vorticity power cascades to smaller
scales (larger $l$) as the flow evolves with time. At later times,
for the non-rotating case ($\mathrm{Ro}=\infty$), the vorticity power
still remain nearly uniformly distributed over all of the comprising
zonal modes, and isotropic with respect
to the zonal modes. With the Rossby number decreasing from infinity
to $2.08 \times 10^{-3}$, the vorticity power distribution tends to become 
 concentrated 
in the smaller zonal modes and the distribution of the vorticity power
becomes anisotropic with respect to the zonal modes. This represents
the zonalization of the vortical structures and transition to Rossby
wave like motions from turbulent flow with decreasing Ro. With further
decrease in Ro, however, all of the comprising zonal modes tend to
become equally significant, i.e., the vorticity power returns to
isotropy in the zonal modes, and this is more clearly observable at late
times. Thus, the zonalization of the vortical structures with decreasing
Ro is non monotonic for the present case.

\begin{figure*}
\centering{}
\subfloat[]{
\centering{}\includegraphics[scale=0.22]{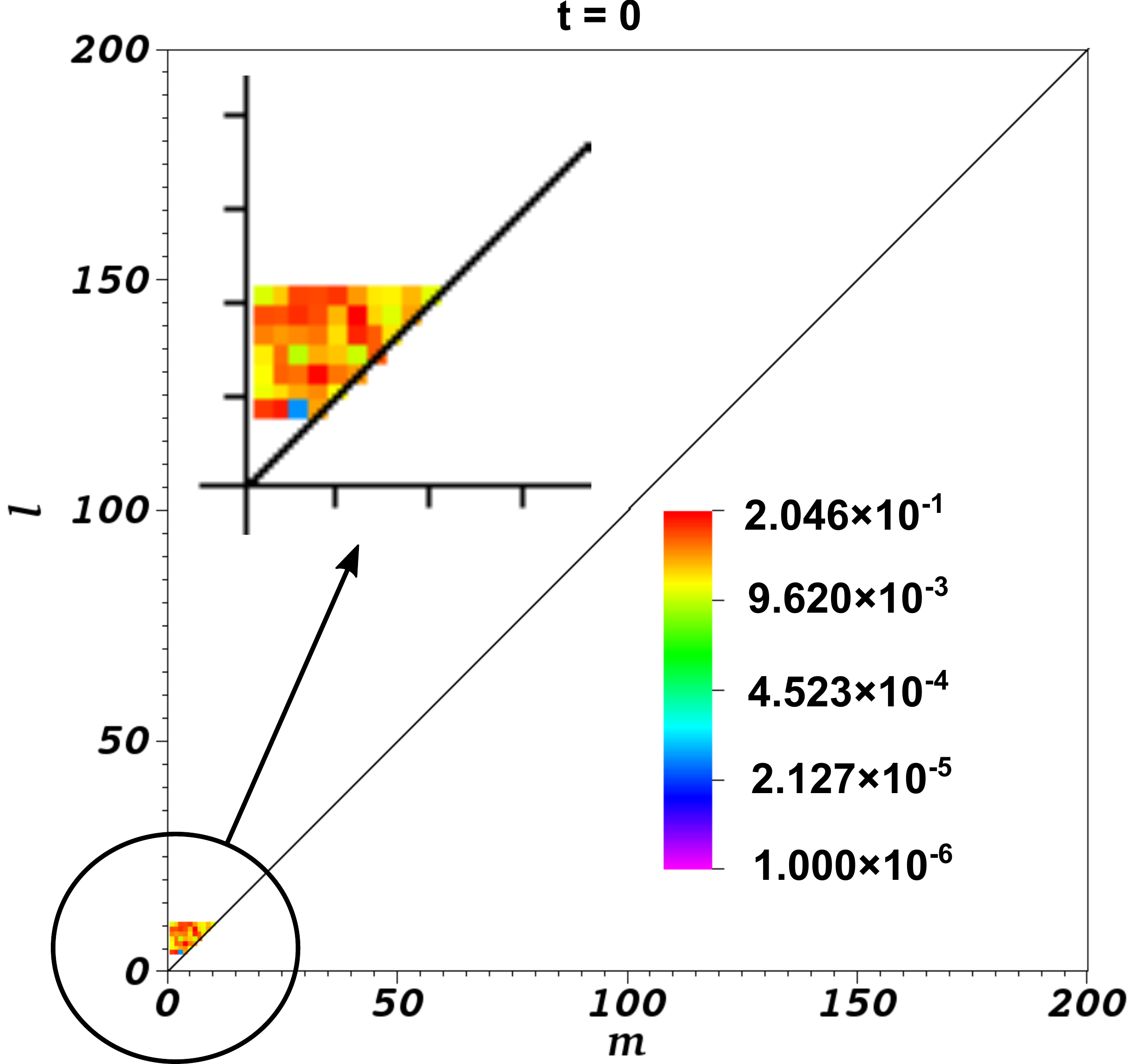}}
\hspace{3mm}\subfloat[]{\centering{}\includegraphics[scale=0.15]{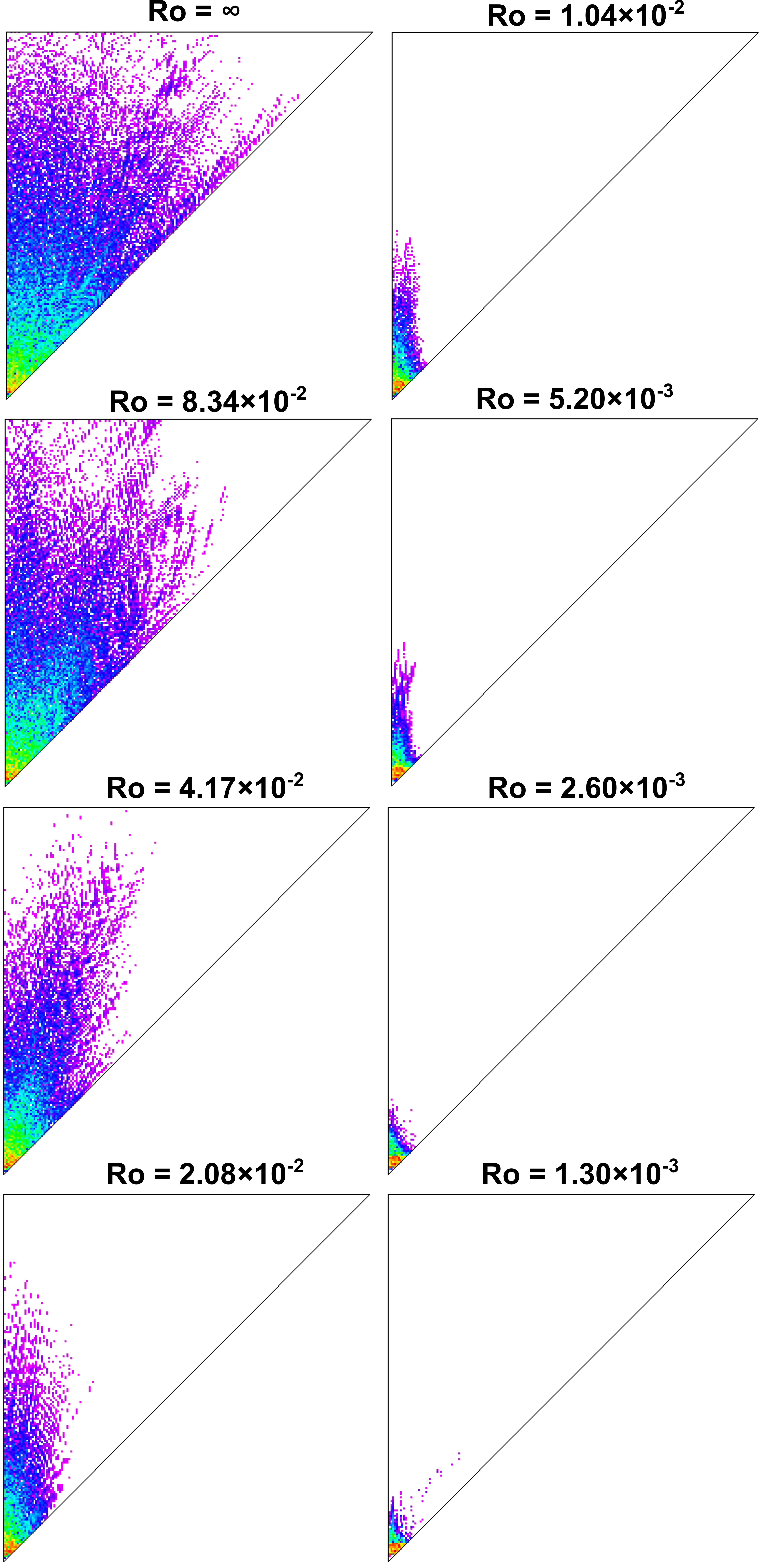}}
\caption{}
\end{figure*}
\begin{figure*}
\ContinuedFloat 
\centering{} 
\subfloat[]{
\centering{}\includegraphics[scale=0.15]{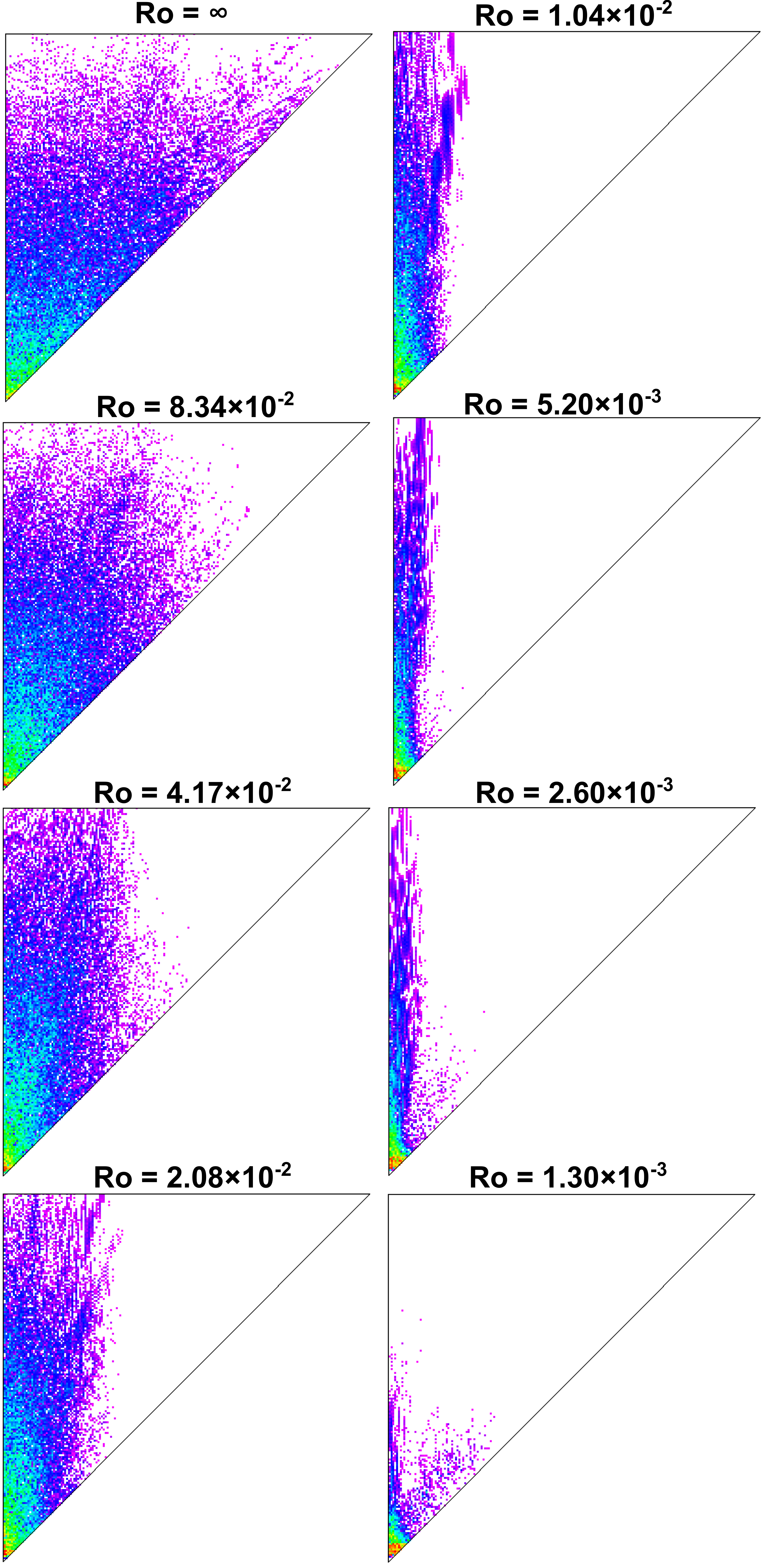}}\hspace{3mm}\subfloat[\label{fig:t-=00003D-92.304}]{
\begin{centering}
\includegraphics[scale=0.15]{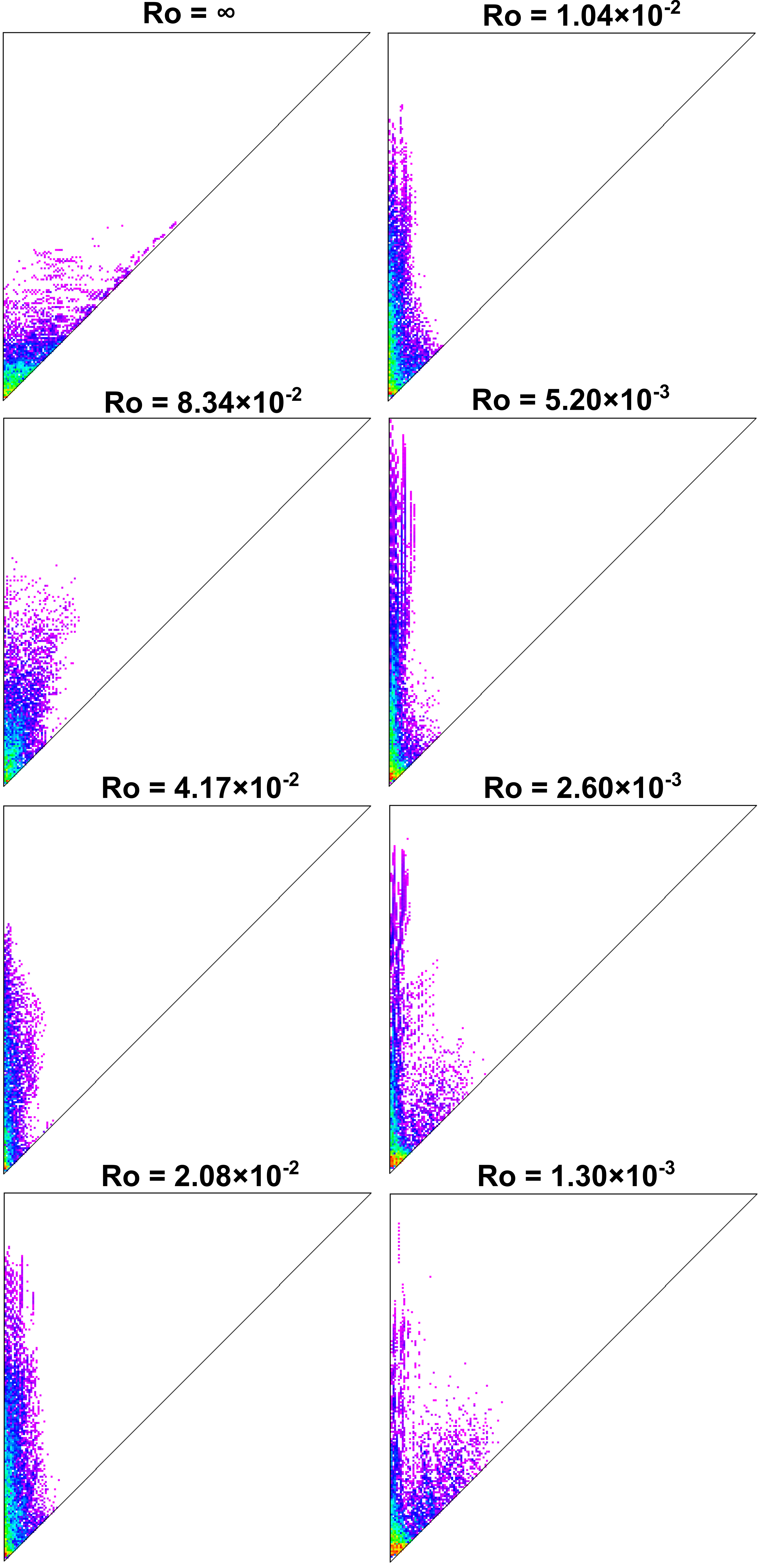}
\par\end{centering}
}\caption{Vorticity power distribution in spectral space ($l,m$) for different Ro, showing diminishing enstrophy cascade with decreasing Ro and zonalization of the power. Time instants shown are (a) $t = 0$, (b) $t = 1.846$, (c) $t = 18.461$, (d) $t = 92.304$ \label{fig:Vorticity-power-distribution}}

\end{figure*}

Table \ref{tab:Rhynes-scale-for} shows the Rhines scales \citep{rhines_1975} for the present case. We choose the reference latitude to be 45{\textdegree} for the computation of $\beta$ here. For Ro$\leq 1.04 \times 10^{-2}$, the wavenumbers ($l=4-10$) comprising the initial vorticity field are smaller than the Rhines scale (as a wavenumber, $k_{\beta}$), i.e., the scales comprising the initial vorticity field are larger than the Rhines scale (as a length scale, 1/$k_{\beta}$). Hence, according to the Rhines theory \citep{rhines_1975}, for the present case for Ro$\leq 1.04 \times 10^{-2}$, the eddies in the initial vorticity field should not merge and initiate the inverse energy cascade (and therefore the forward enstrophy cascade) because the initial scales are already larger than the Rhines scale. However, the spectral analysis (see figures \ref{fig:Vorticity-power-spectra}, \ref{fig:Vorticity-power-distribution}) does show forward enstrophy cascade even for Ro$\leq 1.04 \times 10^{-2}$, indicating that the cascade does not cease completely at the Rhines scale.

\begin{center}
\begin{table}

\begin{centering}
\begin{tabular}{|ccccc|}
%\cline{1-2} \cline{4-5} 
\hline
Ro & $k_{\beta}$ &  & Ro & $k_{\beta}$\tabularnewline
%\cline{1-2} \cline{4-5} 
\hline
8.34$\times10^{-2}$ & 3.878  &  & 5.20$\times10^{-3}$ & 15.514 \tabularnewline
%\cline{1-2} \cline{4-5} 
4.17$\times10^{-2}$ & 5.485 &  & 2.60$\times10^{-3}$ & 21.940\tabularnewline
%\cline{1-2} \cline{4-5} 
2.08$\times10^{-2}$ & 7.757 &  & 1.30$\times10^{-3}$ & 31.027\tabularnewline
%\cline{1-2} \cline{4-5} 
1.04$\times10^{-2}$ & 10.970 & \multicolumn{1}{c}{} & \multicolumn{1}{c}{} & \multicolumn{1}{c|}{}\tabularnewline
%\cline{1-2} 
\hline
\end{tabular}
\par\end{centering}
\caption{Rhines scale for case A\label{tab:Rhynes-scale-for}}
\end{table}
\par\end{center}

\subsection{Vorticity probability density}

The vorticity probability density/vorticity measure is defined as
the area occupied by each vorticity level and expressed as 

\begin{equation}
p\left(\omega\right)=\frac{\textrm{the area occupied by vorticity in the range \ensuremath{\left[\omega+\Delta\omega\right]}}}{4\pi \Delta\omega}. \label{eq:pd_vorticity}
\end{equation}

\begin{figure*}
\begin{centering}
\subfloat[]{
\begin{centering}
\includegraphics[scale=0.35]{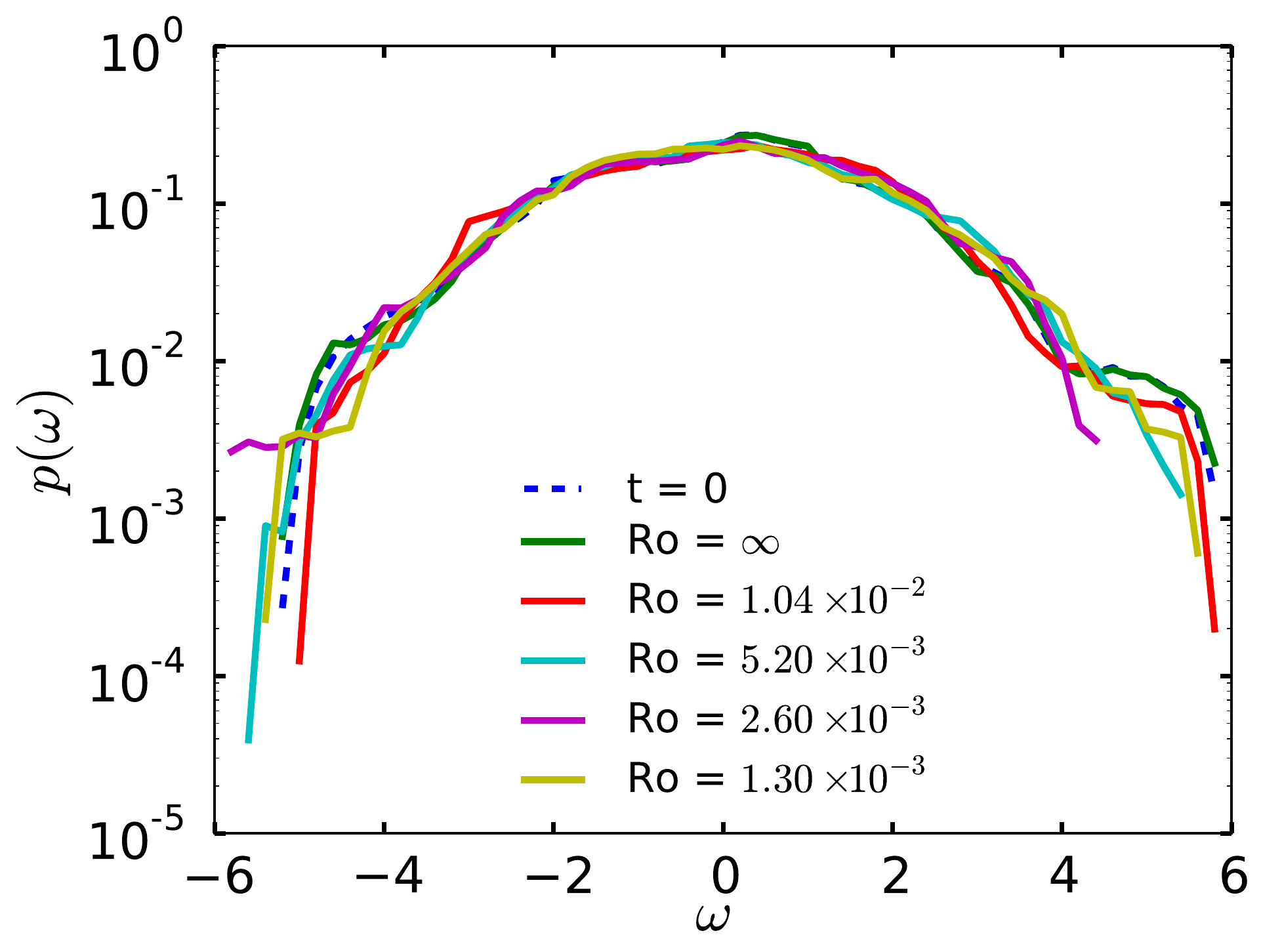}
\par\end{centering}
}\subfloat[]{
\centering{}\includegraphics[scale=0.35]{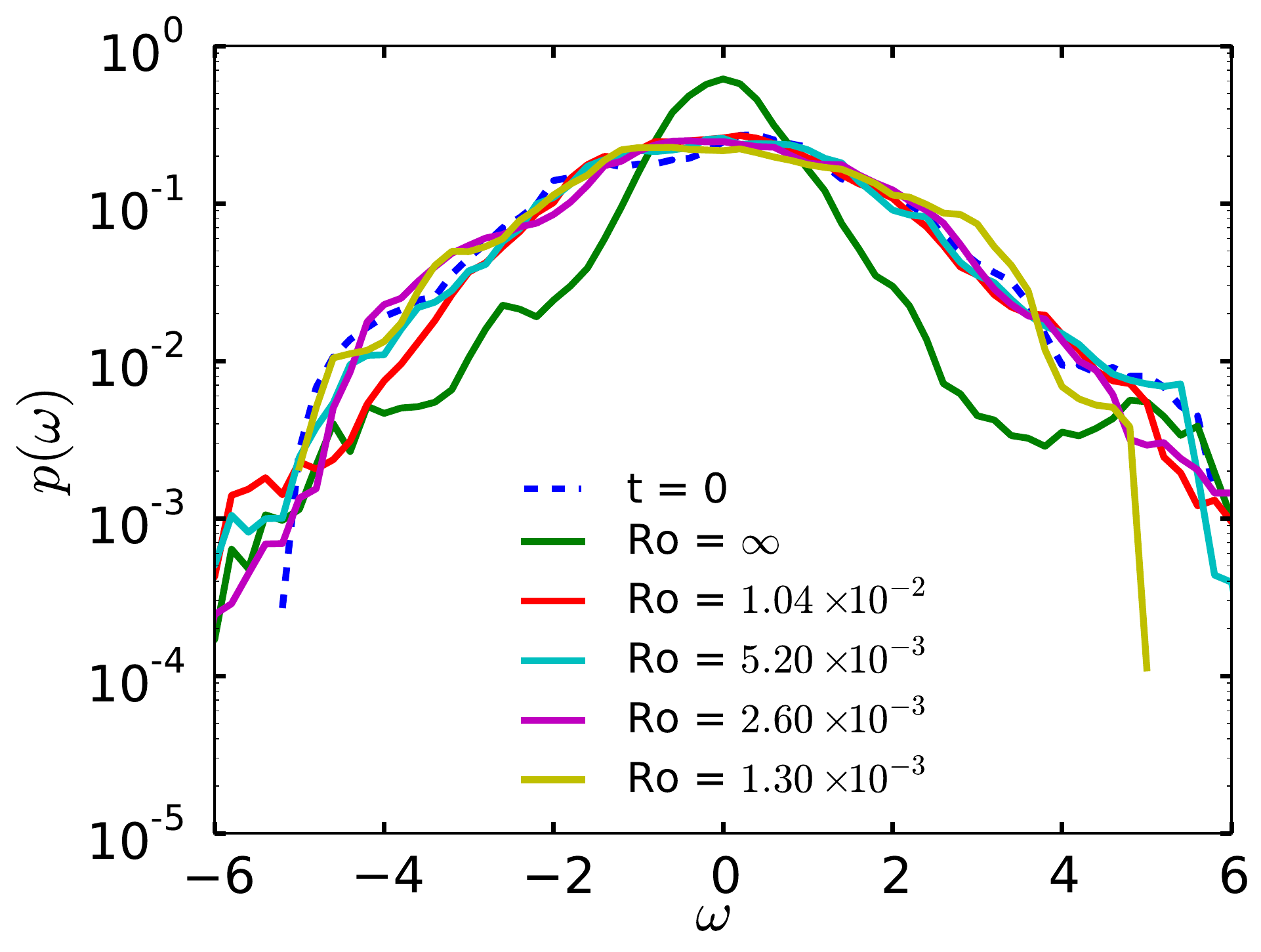}}
\par\end{centering}
\centering{}\subfloat[]{
\begin{centering}
\includegraphics[scale=0.35]{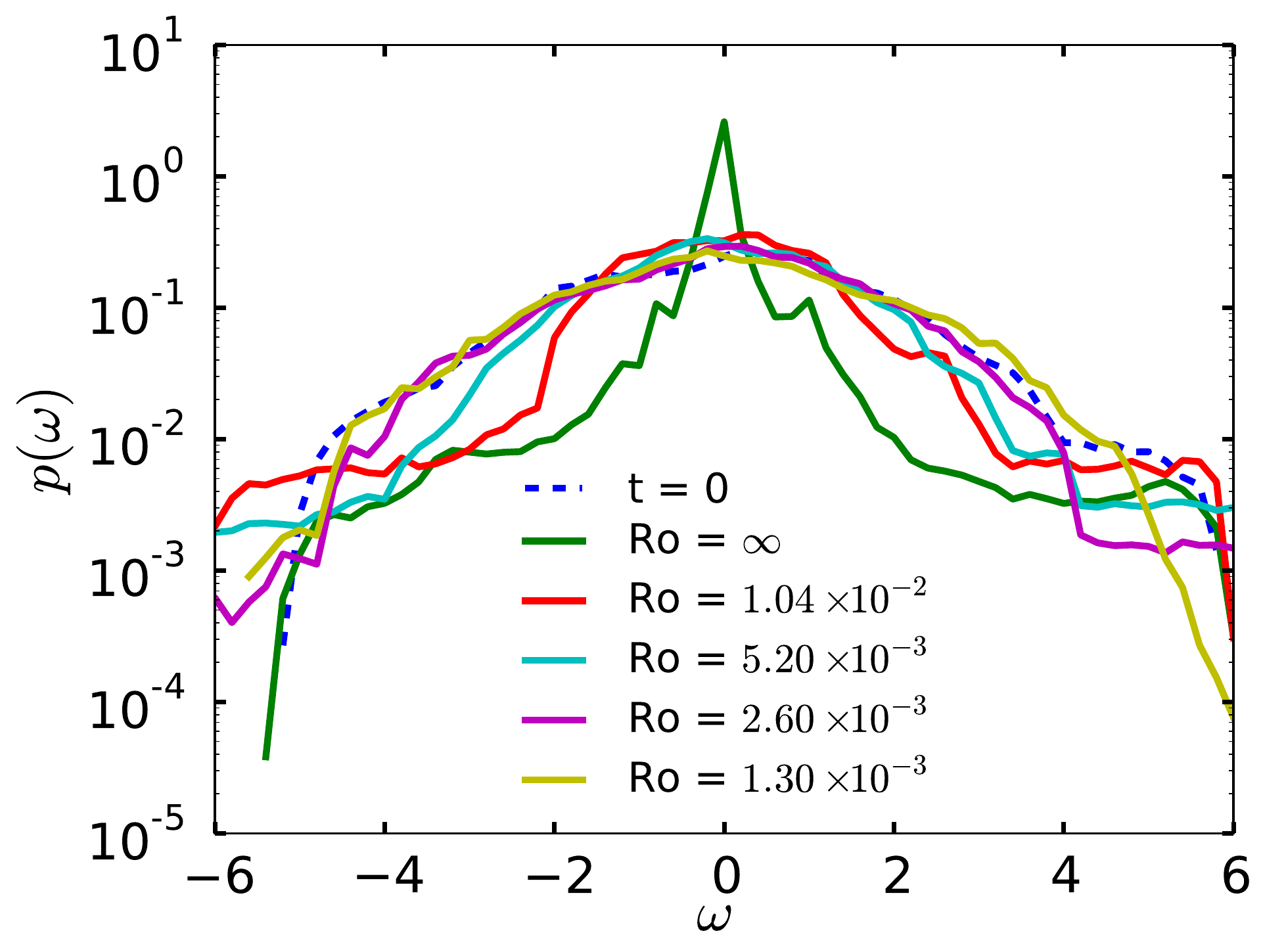}
\par\end{centering}
}\caption{Vorticity probability density as a function of time and Ro. The distribution of probability density becomes broader with decreasing Ro. (a) $t=1.846$, (b) $t=18.461$, (c) $t=92.304$. \label{fig:Vorticity-probability-density}}
\end{figure*}

The conservation of all Casimirs/inviscid invariants is equivalent
to the conservation of $p\left(\omega\right)$ $\left ( p\left(\omega + f \right) \right )$ for the non-rotating (for the rotating) inviscid flow.
However, there is a lack of conservation in simulations at finite
resolution because of the inevitable cascade of vorticity to small
scales \citep{dritschel2015late}. Here, we investigate the probability density of vorticity in the same spirit as Dritschel {\it et al.} \citep{dritschel2015late}. Figure \ref{fig:Vorticity-probability-density}
shows the vorticity probability density for different Rossby numbers
at $t=1.846,\,18.461$ and $92.304$. The initial (at $t=0)$ vorticity
probability density is broadly distributed between $\omega_{min}=-5.1187$
and $\omega_{max}=5.7996$. For the non-rotating case, $p\left(\omega\right)$
narrows and diminishes everywhere except for values of $\omega$ near
zero at late times. However, for the rotating cases, the distribution
of $p\left(\omega\right)$ remains broad even at late times, and smaller
the Rossby number, broader is the distribution of $p\left(\omega\right)$
at late times. This shows that as the Rossby number decreases, the
forward cascade of vorticity to smaller scales decreases, consistent
with the inferences made from the vorticity and power spectra plots.

\subsection{Aspect ratio and orientation of the vortical
structures}

We identify individual vortical structures using an algorithm (see
appendix \ref{appB}). We determine the centroid of each individual structure
from the computation of first moment of area. Hence, we have

\begin{equation}
\bar{\theta}_{i}=\frac{\int_{\Omega}\theta_{i}\left|\omega\right|\textrm{d}\Omega}{\int_{\Omega}\left|\omega\right|\textrm{d}\Omega}=\frac{\sum_{n=1}^{N}\left(\theta_{i}\right)_{n}\left|\omega\right|_{n}\left(\Delta\Omega\right)_{n}}{\sum_{n=1}^{N}\left|\omega\right|_{n}\left(\Delta\Omega\right)_{n}};i=1,2. 
\end{equation}

Here, $\theta_{1}$ and $\theta_{2}$ are the coordinates longitude
and latitude respectively, and the overbar stands for the centroid
coordinates, $\left|\omega\right|$ is the vorticity magnitude, $\Omega$
is the total surface area of the sphere, $\Delta\Omega$ is the dual
area surrounding a primal node, and $N$ is the total number of primal
nodes. We approximate each individual structure by an ellipse, and
determine the major and minor axes from the computation of the second
moment of area as follows. The expression of second moment of area
is as follows.

\begin{eqnarray}
& \mu_{ij}=\frac{\int_{\Omega}\left(\theta_{i}-\bar{\theta}_{i}\right)\left(\theta_{j}-\bar{\theta}_{j}\right)\left|\omega\right|\textrm{d}\Omega}{\int_{\Omega}\left|\omega\right|\textrm{d}\Omega} \nonumber \\ &=\frac{\sum_{n=1}^{N}\left[\left(\theta_{i}\right)_{n}-\bar{\theta}_{i}\right]\left[\left(\theta_{j}\right)_{n}-\bar{\theta}_{j}\right]\left|\omega\right|_{n}\left(\Delta\Omega\right)_{n}}{\sum_{n=1}^{N}\left|\omega\right|_{n}\left(\Delta\Omega\right)_{n}};i=1,2,j=1,2. 
\end{eqnarray}

Then, we compute a covariant matrix as

\begin{equation}
M=\left[\begin{array}{cc} \mu_{11} & \mu_{12}\\ \mu_{21} & \mu_{22} \end{array}\right]. 
\end{equation} 

\begin{figure}
\centering{}\includegraphics[scale=0.35]{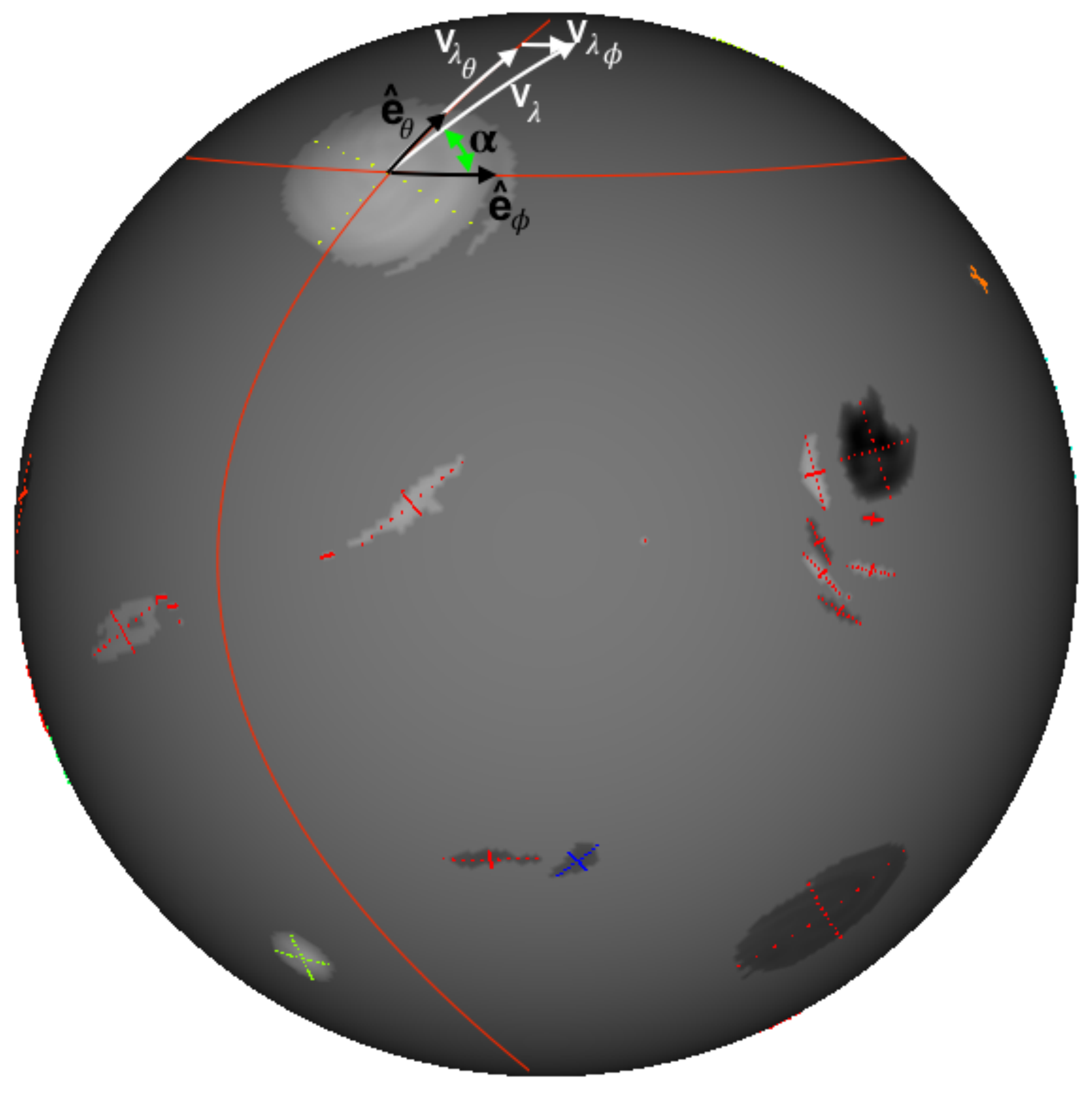}\caption{A representative plot showing individual vortical structures, major and minor axes, and the definition of angle $\alpha$.
\label{fig:vortical_structures}}
\end{figure}

Now, we determine the lengths of the major and minor axes of the vortical structure
from the larger ($\lambda_{larger}$) and smaller ($\lambda_{smaller}$) eigenvalues of the covariant matrix, respectively. The aspect ratio, a measure of the elongation,
of the vortical structure is computed as the ratio of the major axis length to
the minor axis length. The orientation of the vortical structure is
determined from the angle ($\alpha$, see the schematic in figure \ref{fig:vortical_structures}) between
the major axis and the azimuthal direction. We determine this angle from the dot product of
the eigenvector corresponding to the larger eigenvalue with the
basis vector in the azimuthal direction. A representative plot showing individual structures, the major and minor axes of these structures, and the definition of the angle $\alpha$ is given in figure \ref{fig:vortical_structures}. We compute the probability density of the aspect ratio or $\cos\alpha$ as

\begin{equation}
p\left(f\right)=\frac{\textrm{The number of \ensuremath{f} values in the range \ensuremath{\left[f+\Delta f\right]}}}{\textrm{Total number of \ensuremath{f} values \ensuremath{\times}}\Delta f}.
\end{equation}

\begin{figure}

\centering{}\includegraphics[scale=0.4]{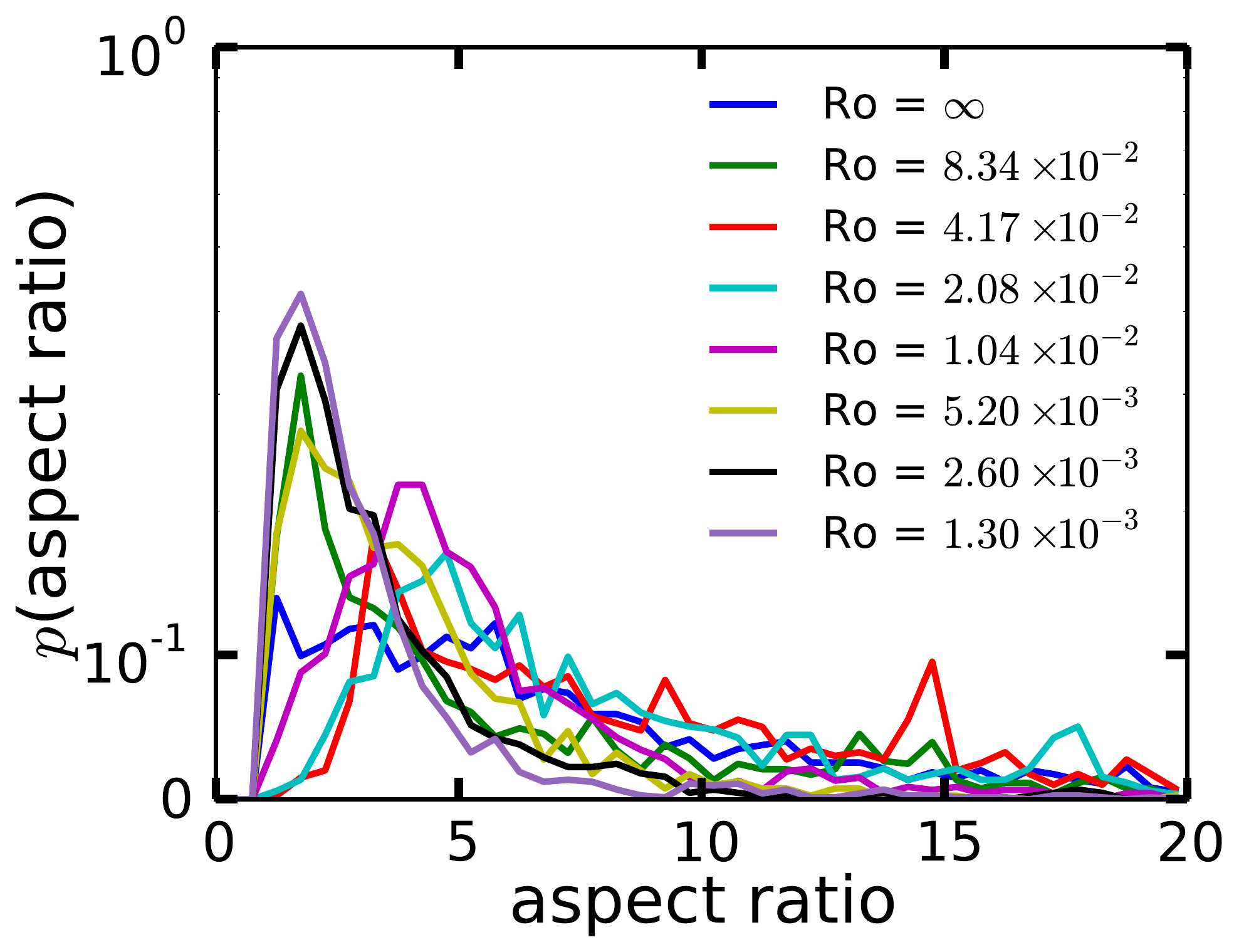}\caption{Probability density function of the aspect ratio of individual vortices
as a function of Ro over the turnover time from 69.228 - 115.379.
\label{fig:Probability-density-function-aspect-ratio}}
\end{figure}

\begin{figure}
\centering{}\includegraphics[scale=0.4]{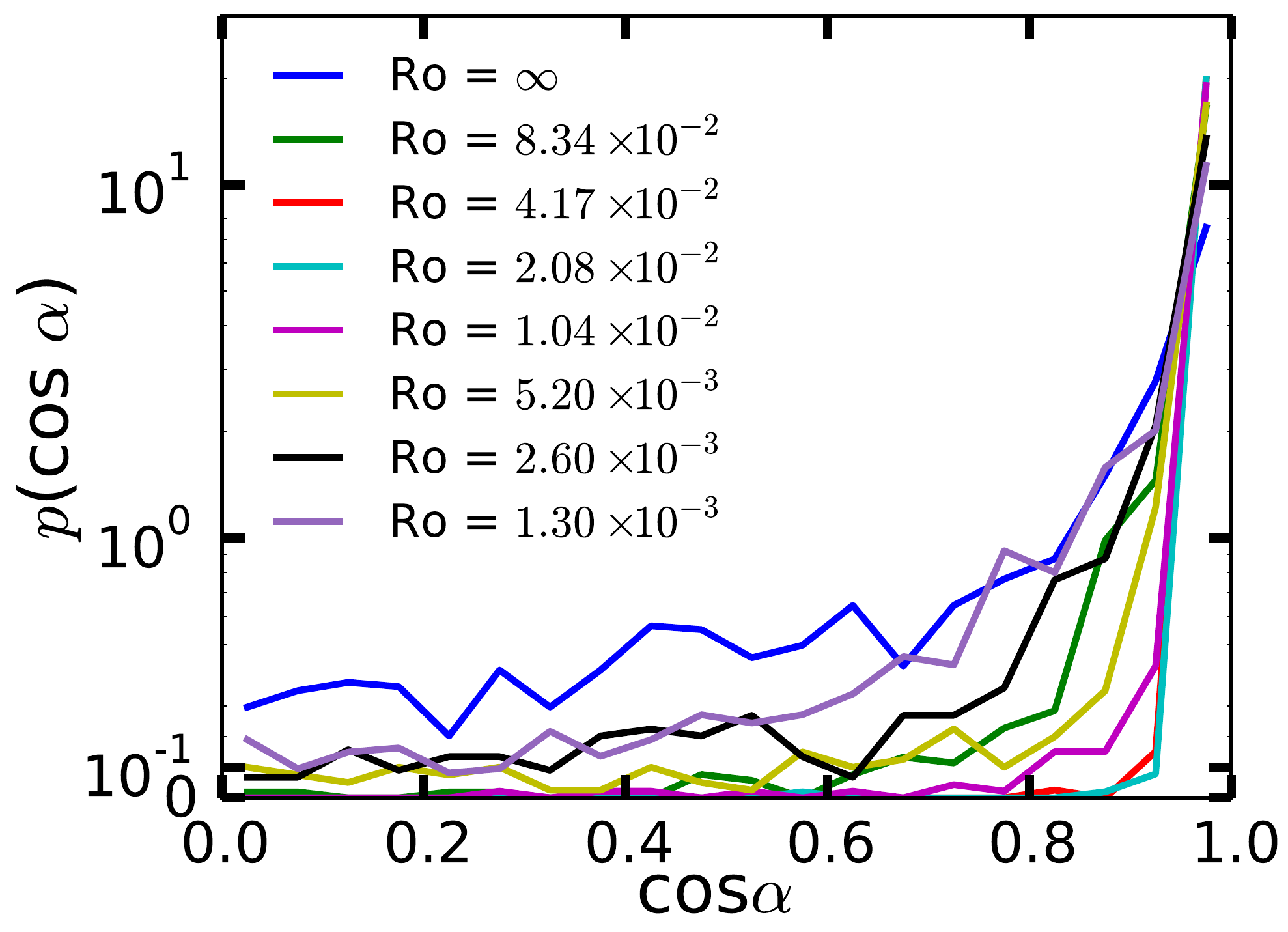}\caption{Probability density function of the cosine of the angle between the
major axis and the azimuthal direction as a function of Ro over the
turnover time from 69.228 - 115.379.\label{fig:Probability-density-function-orientation}}
\end{figure}

Here, the function $f$ represents the aspect ratio or $\cos\alpha$ distribution. Figure \ref{fig:Probability-density-function-aspect-ratio} shows
the probability density of the aspect ratio of the vortical structures
as a function of Ro at late times (over the turnover time from 69.228
- 115.379). Here, the structures with aspect ratio > 20 are excluded for 
 the sake of clarity. The lower limit of the aspect ratio (= 1) represents circular structures, and the higher values represent the elongated structures. 
For the non-rotating case (Ro$=\infty$), the
peak in the value of probability density at aspect ratio $\approx1$
corresponds to the presence of quadrupolar vorticity field. The decreasing
probability density at larger aspect ratios corresponds to the presence
of small scale vortices due to the forward enstrophy cascade. With
the Rossby number decreasing from infinity to $2.08 \times 10^{-2}$, the probability
density peak moves away from aspect ratio $\approx1$, showing that
the vortical structures tend to elongate with decreasing Ro. In fact,
these elongated structures tend to be zonal (see figure \ref{fig:Probability-density-function-orientation}).
With further decrease in Ro, the probability density peak tends back to
move towards aspect ratio $\approx1$. This shows that the vortical
structures tend back to become circular and non-zonal (see figure \ref{fig:Probability-density-function-orientation})
with further decrease in Ro. This is consistent with the aforementioned
claim that the zonalization of the vortical structure is non-monotonic
with decreasing Ro (for the present case).

Similarly, figure \ref{fig:Probability-density-function-orientation}
shows the probability density of the orientation of the vortical structures
(that of the cosine of the angle between the major axis and the azimuthal
direction) as a function of Ro at late times (over the turnover time
from 69.228 - 115.379). The upper limit $\cos\alpha=1$ represents the structures aligned in the zonal direction, and the lower values represent the non-zonal structures. For the non-rotating case (Ro$=\infty$), the relatively higher values of $p\left(\cos\alpha\right)$ at larger $\cos\alpha$
values correspond to the presence of small scale vortices due to the
forward enstrophy cascade. For the rotating cases,
with Ro decreasing up to
$2.08 \times 10^{-2}$ from infinity, $p\left(\cos\alpha\right)$ tends to increase
at $\cos\alpha\approx1$, and to vanish at smaller values of $\cos\alpha$.
This indicates the zonalization of the structures with decreasing
Ro. Moreover, these zonal structures have larger aspect ratio / are
elongated (see figure \ref{fig:Probability-density-function-aspect-ratio}).
With further decrease in Ro, $p\left(\cos\alpha\right)$ tends to decrease
at $\cos\alpha\approx1$, and to augment at smaller values of $\cos\alpha$.
Thus, the structures tend to become non-zonal and circular (see figure
\ref{fig:Probability-density-function-aspect-ratio}, the aspect ratio
of these structures tend to be smaller) with further decrease in Ro.
This further confirms the aforementioned claim of non-monotonic nature
of the zonalization of the vortical structure with decreasing Ro (for the present case). 

\subsection{Effect of wavenumber rage constituting the (arbitrary) initial vorticity field}
We consider two different arbitrary initial vorticity fields comprising of wavenumbers $l=4-10$ (test cases B, and C), and three arbitrary initial vorticity fields comprising of wavenumbers $l=4-20$, $4-40$ and $4-80$ (test cases D, E, and F). We investigate the effect of these initial conditions on the evolution of the vorticity field. The results are presented in the following sub-sections.

\subsubsection{Influence of initial vorticity distribution}
\begin{figure*}
\centering{}\includegraphics[scale=0.125]{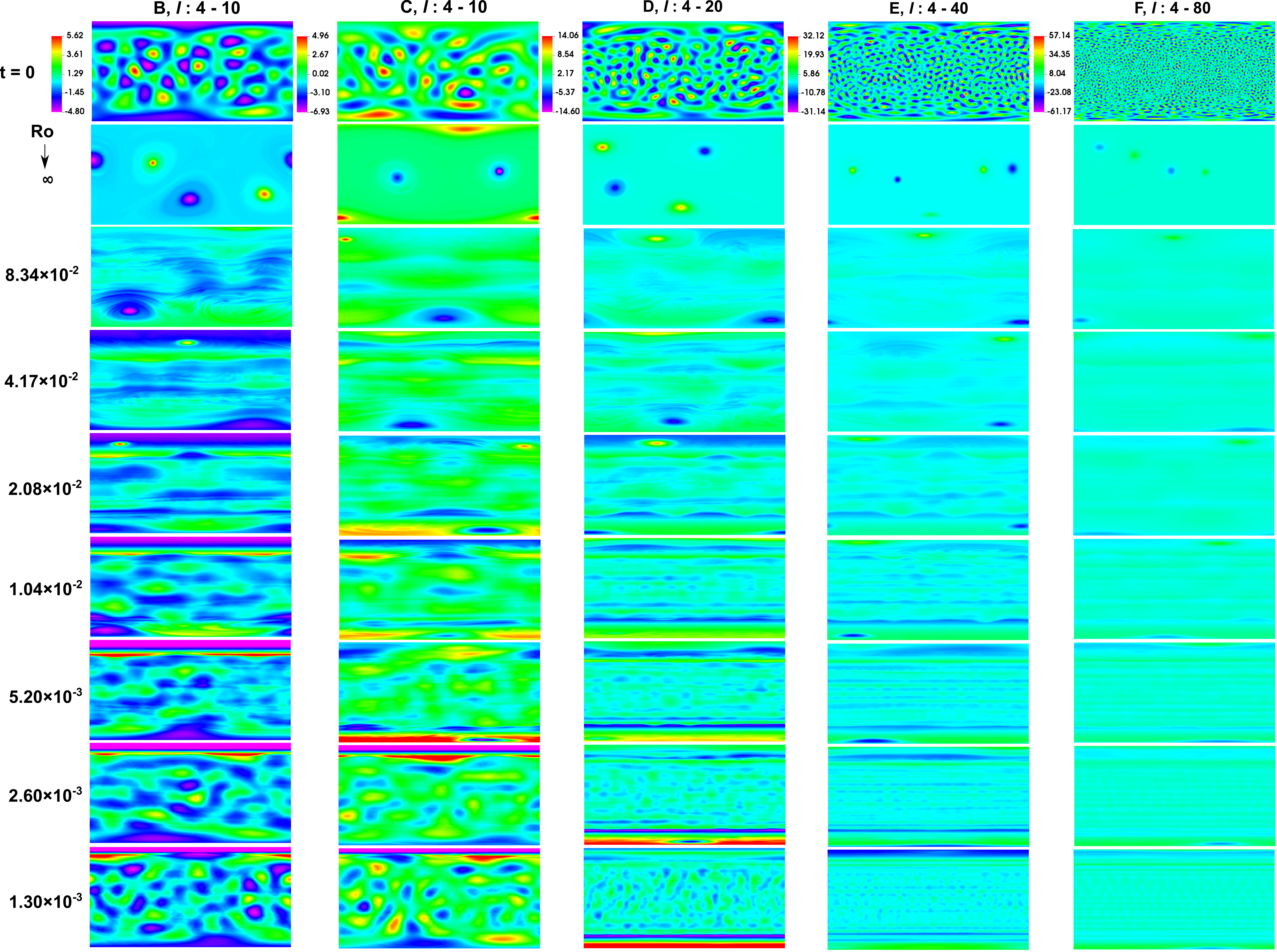}\caption{Vorticity distribution showing the effect of wavenumber rage comprising the (arbitrary) initial vorticity field.\label{fig:Effect-of-wavenumber-range}}
\end{figure*}

So far we have examined the details of vorticity evolution for case A wherein the initial vorticity was confined to the wavenumber range $l=4-10$. 
Figure \ref{fig:Effect-of-wavenumber-range} shows the evolution of the vorticity field for different initial wavenumber distributions. For each distribution we ensure that the total kinetic energy is approximately constant. Although, the initial wavenumber range for the cases B and C is the same (also, it is the same for case A), the initial amplitudes are different for these cases.   
For the initial vorticity field comprising of intermediate wavenumbers ($l=4-10$, test cases B, and C), at a later time, the emerging vortices from the merger of smaller scales (due to inverse cascade) tend to be zonal due to the rotation. However, the zonalization is non-monotonic with decreasing Ro. Whereas, for the initial vorticity field comprising of intermediate to large wavenumbers ($l=4-20$, $l=4-40$, and $l=4-80$; test cases D, E, and F),  
the zonalization tends to be monotonic with decreasing Ro. Moreover, the enstrophy cascade diminishes but does not completely cease with decreasing Ro even for the initial vorticity spectrum comprising of the scales larger than the Rhines scale. This diminishing effect tends to be weaker from test case B to F. As we move from the test case B to F, the range of scales smaller than the Rhines scale present in the initial vorticity field increases. Therefore, the cascade becomes stronger from test case B to F, and therefore the effect of rotation becomes weaker from B to F. Moreover, the scales formed from the merger of smaller scales during the inverse cascade tend to be zonal. Hence, the stronger cascade from B to F 
 is a probable cause for the zonalization to become non-monotonic to monotonic.   

\subsubsection{Spectral distribution of vorticity power}
\begin{figure*}
\centering{}\includegraphics[scale=0.12]{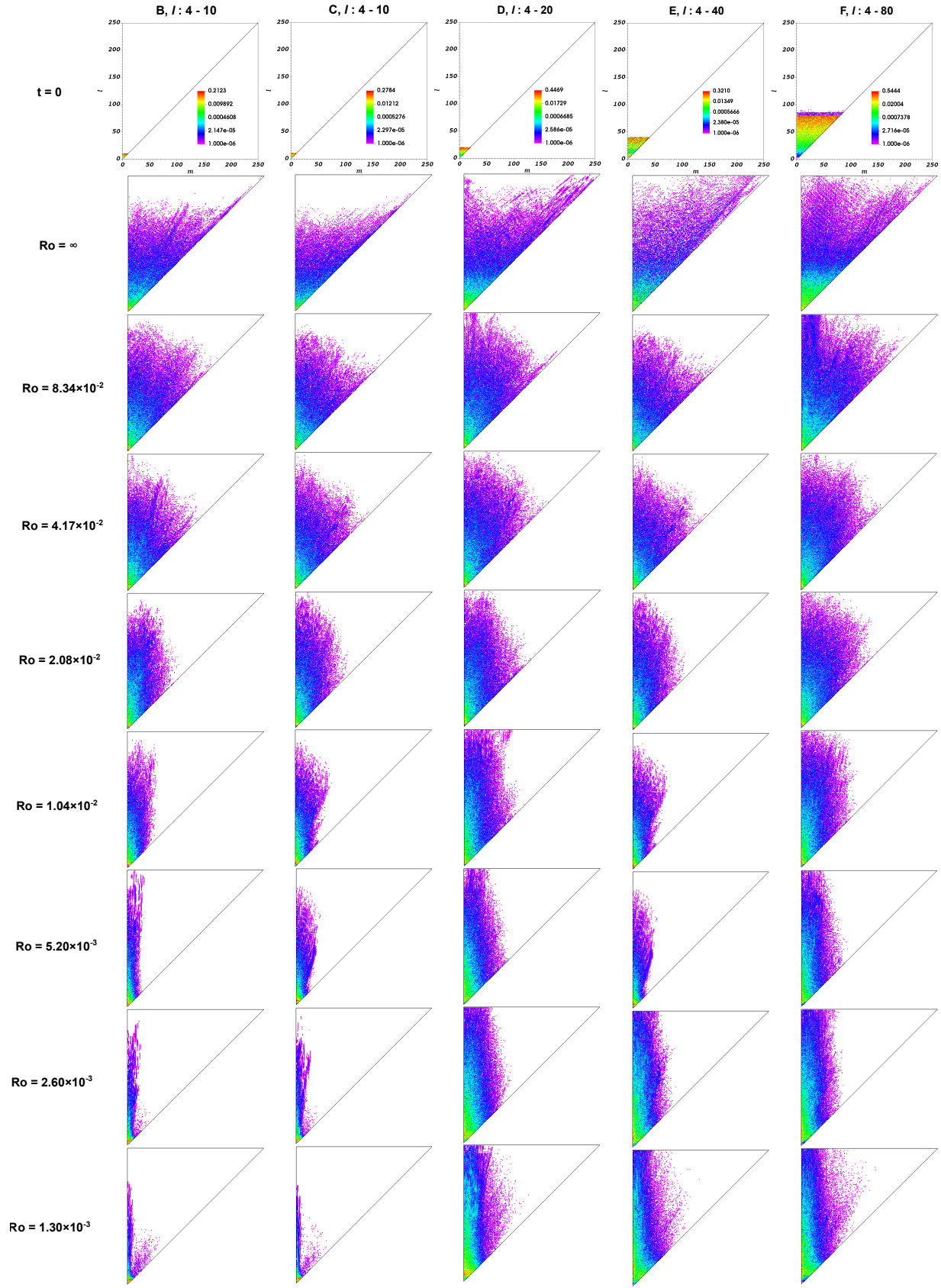}\caption{Spectral distribution of vorticity power showing the effect of wavenumber rage comprising the (arbitrary) initial vorticity field.\label{fig:Vorticity-power-Effect-of-wavenumber-range}}
\end{figure*}

Figure \ref{fig:Vorticity-power-Effect-of-wavenumber-range} shows the distribution of vorticity power in the spectral $\left( m - l \right)$ space as a function of the wavenumber range constituting the initial  vorticity field (test cases B - F, see table \ref{tab:Simulation-parameters})  and Ro. For cases B and C, at a later time ($t\approx18$), as Ro decreases the vorticity power tends to confine to smaller $m$, i.e., there is zonalization of the vortical structures. However, for  Ro smaller than $5.20 \times 10^{-3}$, the vorticity power tends also to be equally significant for larger $m$, showing non-monotonic nature of the zonalization. However, the confinement of the vorticity power to smaller $m$ with decreasing Ro tends to be monotonic for the cases D, E, and F, i.e., the zonalization tends to be monotonic with decreasing Ro for these cases. 
 
\subsubsection{Probability density of orientation of vortical structures}
\begin{figure*}
\centering{}\includegraphics[scale=0.25]{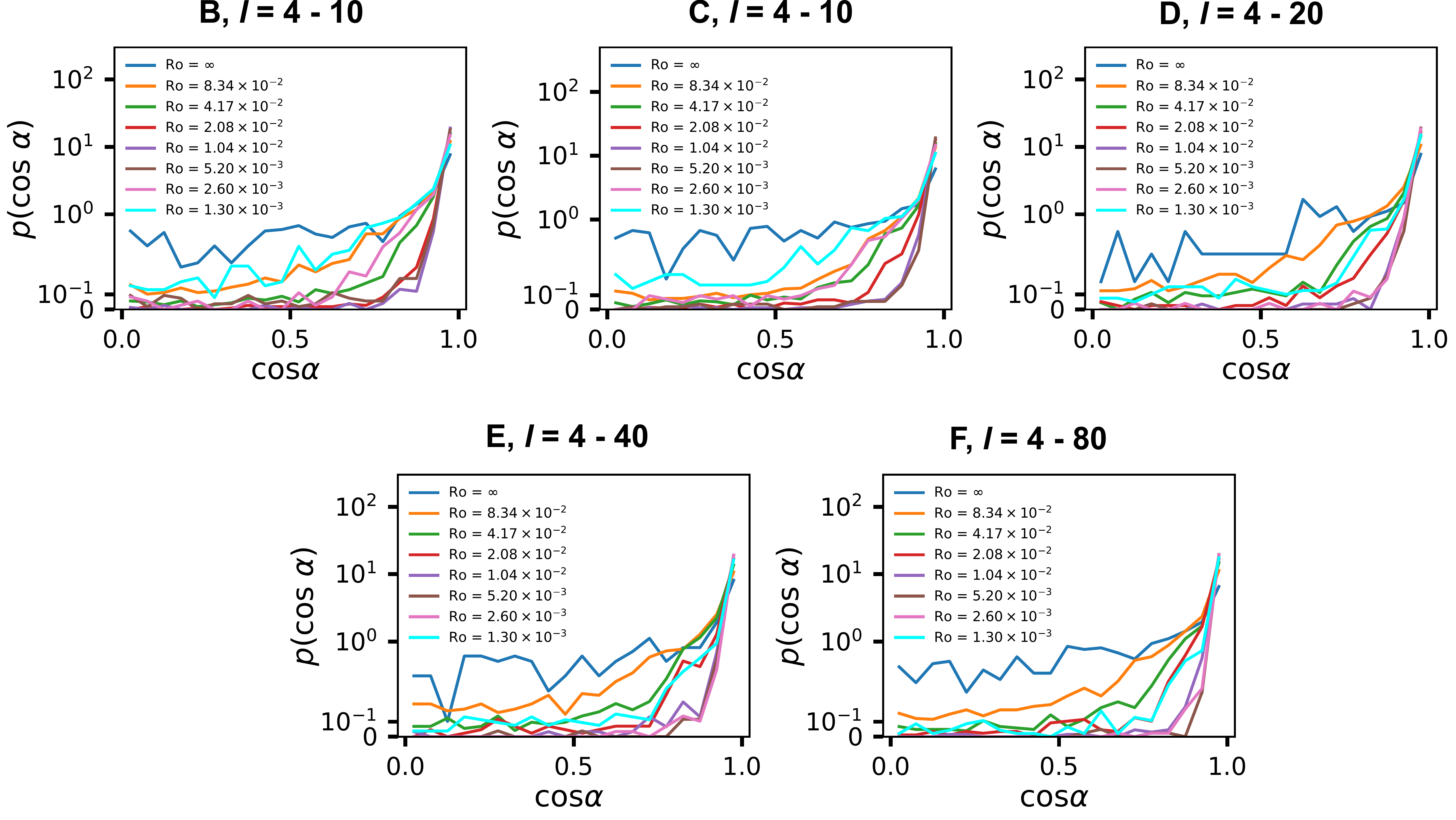}\caption{Probability density of cos$\alpha$ showing the effect of wavenumber rage comprising the (arbitrary) initial vorticity field.\label{fig:pdf-cos-alpha-Effect-of-wavenumber-range}}
\end{figure*}

Figure \ref{fig:pdf-cos-alpha-Effect-of-wavenumber-range} shows the probability density of cosine of the angle between the major axis of the vortical structure and the azimuthal direction (that of the orientation of the vortical structures) as a function of the wavenumber range constituting the initial  vorticity field and Ro. For the intermediate wavenumber range comprising the initial vorticity field (for cases B, and C), as the Ro decreases, the probability density diminishes for all of the cos$\alpha$ values except near one up to Ro of about $5.20 \times 10^{-3}$. This shows the zonalization of the vortical structures. With further reduction in Ro, the probability density for cos$\alpha$ values smaller than one tends back to augment, showing nonmonotonic nature of the zonalization with decreasing Ro. However, for the initial vorticity field comprising of intermediate to large wavenumbers (for cases D, E, and F), the diminishing of the probability density for all of the cos$\alpha$ values except near unity is almost monotonic with decreasing Ro (there is insignificant augmentation of probability density values at very low Ro), revealing monotonic zonalization.

\section{Conclusions}

The present study examines the effect of rotation on the vorticity dynamics on a unit sphere using discrete exterior calculus. 
In addition to examining the effect of rotation, we also investigate different initial spectra constituting the initial vorticity field and the differences in the late time evolution of 
the vorticity field. 
The visualization of the evolving vorticity field, reveals diminishing of the forward enstrophy cascade and the non-monotonic nature of the zonalization of the vortical structures for the initial vorticity field comprising of intermediate-wavenumbers. On the other hand,  the zonalization is monotonic for the initial vorticity field comprising of intermediate -to- large wavenumbers.  We analyze the vorticity field on the sphere by
computing the vorticity power spectrum,
distribution of spectral power, vorticity probability density, probability
density of aspect ratio and that of the orientation of the vortical
structures. The analyses further confirms the phenomenon of zonalization although we note the tendency to zonalization reverses for high rotation rates where the initial vorticity is confined to a wavenumber range $l=4-10$. Moreover, we observe a forward cascade of enstrophy for the cases with initial vorticity field comprising of larger scales than the Rhines scale. Thus, while the forward
enstrophy cascade diminishes with decreasing Rossby number,
the cascade does not cease completely. The cases with initial vorticity field comprising of the scales much smaller (larger wavenumbers) than the Rhines scale  have a higher potential for the cascade, and therefore, the diminishing effect of rotation is weaker for these cases. Moreover, the scales emerging from the merger of smaller scales during the inverse cascade tend to be zonal. Hence, the zonalization tend to be monotonic for these cases as the diminishing effect of rotation is weaker. Our investigation suggests that in addition to the dependence of the Rhines scale on $\beta$ and RMS velocity, the dependence on the spectrum constituting the initial vorticity field should be included in order for it to represent the scale arresting the cascade more effectively.

\begin{acknowledgments}
This research was supported by the KAUST Office of Sponsored Research under Award URF/1/3723-01-01.
\end{acknowledgments}

\appendix

\section{Physical Setup and Numerical Procedure}\label{sec:num_proc}
We consider evolving flow on a unit sphere, and the computational
domain consists of a spherical surface of unit radius. We use the DEC procedure of Jagad {\it et al.}\cite{jagad2020primitive}. A brief description of the procedure is as follows. The governing equations consist of Euler equations in a rotating frame of reference.  For an inviscid incompressible flow of a homogeneous fluid with unit density, on a compact smooth Riemannian surface, the vector calculus notation of Euler equations in a rotating frame of reference are as follows 
\begin{equation} 
\frac{\partial\mathbf{v}}{\partial t}+\nabla_{\mathbf{v}}\mathbf{v}+\textrm{grad}_{S}p +f\hat{k}\times\mathbf{v} =0,\label{eq:1-1} 
\end{equation}
\begin{equation} 
\textrm{div}_{S}\mathbf{v}=0,\label{eq:2-1} 
\end{equation} 
\noindent where $\mathbf{v}$ is the tangential surface velocity, $p$ is the effective surface pressure (which includes the centrifugal force), $\nabla_{\mathbf{v}}$ is the covariant directional derivative, $\textrm{grad}_{S}$ is the surface gradient, $\textrm{div}_{S}$ is the surface divergence, $f$ is the Coriolis parameter, and $\hat{k}$ is the unit vector in the direction of the axis of rotation.  For a flavour on scaling, the nondimensional form of equation (\ref{eq:1-1}) can be written as

\begin{equation} 
\frac{\partial\mathbf{v}^{\ast}}{\partial t^{\ast}}+\nabla_{\mathbf{v}}^{\ast}\mathbf{v}^{\ast}+\textrm{grad}_{S}^{\ast}p^{\ast} +\frac{1}{\mathrm{Ro}}\hat{k}\times\mathbf{v}^{\ast} =0.\label{eq:1-1-1} 
\end{equation}

Here, advective time scale is used as the characteristic time scale. Equation (\ref{eq:1-1-1}) shows that for low Ro (high rotation rate), the Coriolis force term dominates over the nonlinear advection term. The 2D coordinate invariant form of equations (\ref{eq:1-1}) - (\ref{eq:2-1}) read \\
\begin{equation} 
\frac{\partial\mathbf{u}}{\partial t} +\ast\left(\mathbf{u}\wedge\ast d\mathbf{u}\right)+dp^{d} + \ast\left(\mathbf{u}\wedge\ast f_{2}\right)=0.\label{eq:8} 
\end{equation}
\begin{equation} 
\ast\textrm{d}\ast u =0.\label{eq10} 
\end{equation}
Here, $ u $ is the 1-form velocity, $\textrm{d}$ is the exterior derivative, $\ast$ is the Hodge star operator,     $\wedge$ is the wedge product operator,     $p^{d}=p+\frac{1}{2}\left( u \left(\mathbf{v}\right)\right)$ is the effective dynamic pressure 0-form, and the Coriolis force $f\hat{k}\times\mathbf{v}=\ast\left( u \wedge\ast f_{2}\right)$, where $f_{2}$ is the 2-form corresponding to $f$.  We consider domain discretization with a primal simplicial mesh and there is a corresponding dual mesh (we consider the circumcentric dual). The discrete forms are now the integral quantities on the mesh elements. The discrete exterior calculus expressions of the governing equation are then expressed as 

\begin{equation} 
\left[\left(-\frac{1}{\Delta t}\right)I-\frac{1}{2}\left(W_{V}\right)^{n+1}\ast_{0}^{-1}\left[-d_{0}^{T}\right]\ast_{1}\right]\left(U^{*}\right)^{n+1}+\ast_{1}^{-1}d_{1}^{T}\left(P^{d}\right)^{n+1}=F\label{eq:energy_preserving}, 
\end{equation}  
\noindent with 
\begin{eqnarray} &F=\left(-\frac{1}{\Delta t}\right)\left(U^{*}\right)^{n}+\frac{1}{2}\left(W_{V}\right)^{n+1}\ast_{0}^{-1}d_{b}\left(V\right)^{n+1}+\frac{1}{2}\left(W_{V}\right)^{n}\ast_{0}^{-1}\left(\left[-d_{0}^{T}\right]\ast_{1}\left(U^{*}\right)^{n}+d_{b}\left(V\right)^{n}\right) \nonumber\\& +\frac{1}{2}\left[\left(W_{V}\right)^{n+1}+\left(W_{V}\right)^{n}\right]\ast_{0}^{-1}f_{dual2}, 
\end{eqnarray} 
\begin{equation} \left[d_{1}\right]\left(U^{*}\right)^{n+1}+\left[0\right]\left(P^{d}\right)^{n+1}=0\label{eq:5}, 
\end{equation}  
\noindent where $U^{\ast}$ is the vector containing mass flux primal 1-form for all mesh primal edges, $V$ is the vector containing the discrete primal velocity 1-forms for all mesh primal edges, and $P^{d}$ is the vector containing discrete dynamic pressure 0-forms for all mesh dual vertices, $\Delta t$ is the discrete time interval, $d_{0}$, $\left[-d_{0}^{T}\right]$, $d_{1}^{T}$ are the discrete exterior derivative operators, $\ast_{0}^{-1}$, $\ast_{1}$, $\ast_{1}^{-1}$ are discrete Hodge star operators. The dual-2 form $f_{dual2}$ corresponds to the Coriolis parameter $f=2\Omega\cos\theta$, where $\Omega$ is the rate of rotation of the sphere, and $\theta$ is the colatitude. The present discretization uses the energy-preserving time integration \citep{mullen2009energy}. We employ a biharmonic viscosity as a means to filter and dissipate very high wavenumber content. The spherical surface domain is discretized using about 0.25M nearly uniform triangles and about 0.13M vertices for all simulation cases considered presently.The set of nonlinear equations represented by equations (\ref{eq:energy_preserving}) - (\ref{eq:5}) are solved using Picard's iterative method for the mass flux 1-form and dynamic pressure degrees of freedom. The convergence criterion of $L_{2}$ norm on the residual $\left\Vert R\right\Vert _{2}\leq10^{-8}$ is used.

\section{Verification of the Numerical Procedure}\label{appB0}
\begin{figure*}
\centering{}
\subfloat[]{\begin{centering}
\includegraphics[scale=0.4]{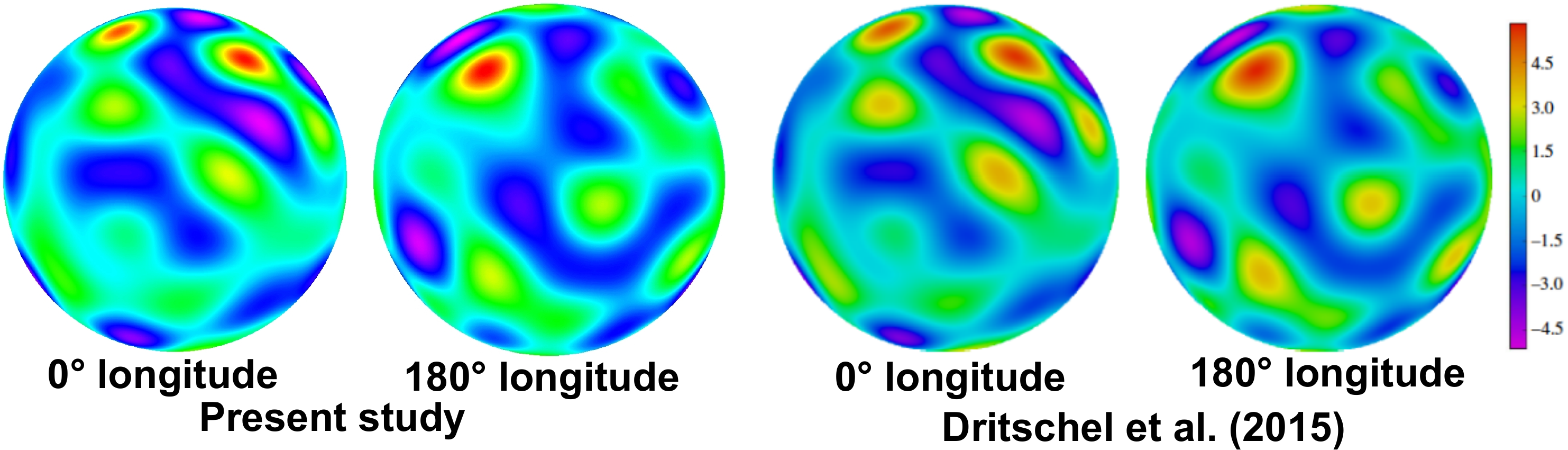}
\label{fig:ic_present_plus_dritschel}
\par\end{centering}
} \\
\subfloat[]{\begin{centering}
\includegraphics[scale=0.4]{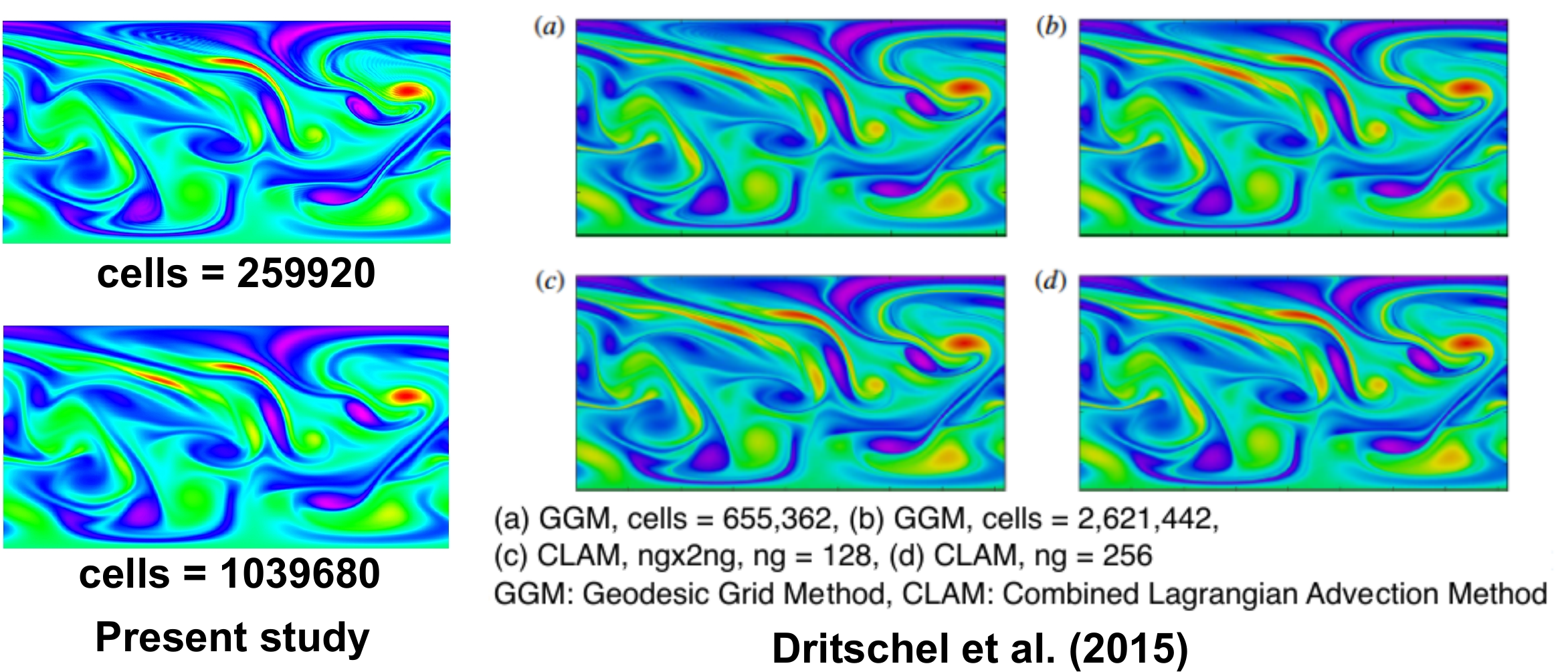}
\label{fig:t_4_present_plus_dritschel}
\par\end{centering}
} \\ \subfloat[]{
\centering{}\includegraphics[scale=0.4]{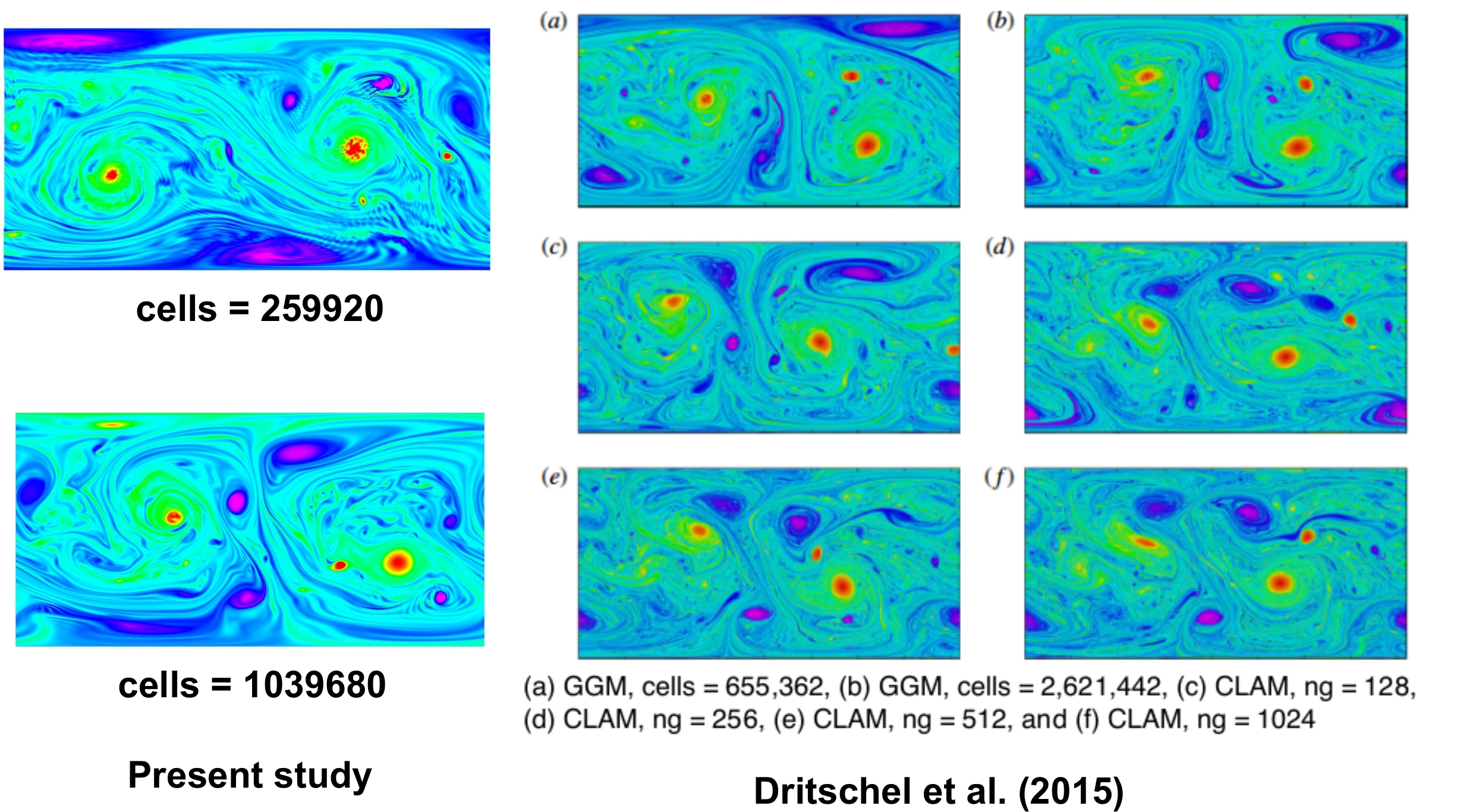} \label{fig:t_40_present_plus_dritschel}}  \\
\subfloat[]{
\centering{} \includegraphics[scale=0.4]{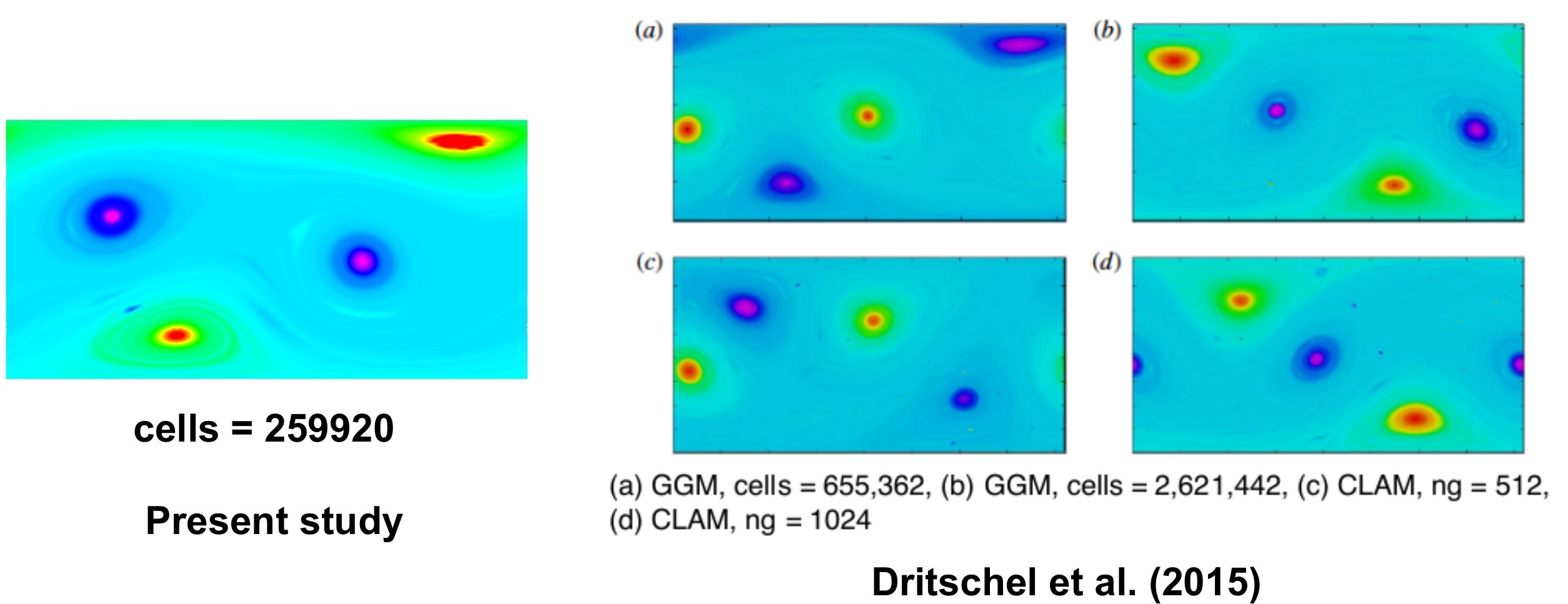} \label{fig:t_400_present_plus_dritschel}} 
\caption{Evolution of the vorticity field from an initial arbitrary vorticity field for the verification case. (a) $t=0$, (b) $t=1.846$, (c) $t=18.461$, (d) $t=184.61$. The right panel figures: Dritschel, D. G., Qi, W., \& Marston, J. B., On the late-time behaviour of a bounded, inviscid two-dimensional flow, Journal of Fluid Mechanics, 783, page 6-9, reproduced with permission.  \label{fig:later_t_present_plus_dritschel}}
\end{figure*}

\begin{figure}
\centering{}\includegraphics[scale=0.4]{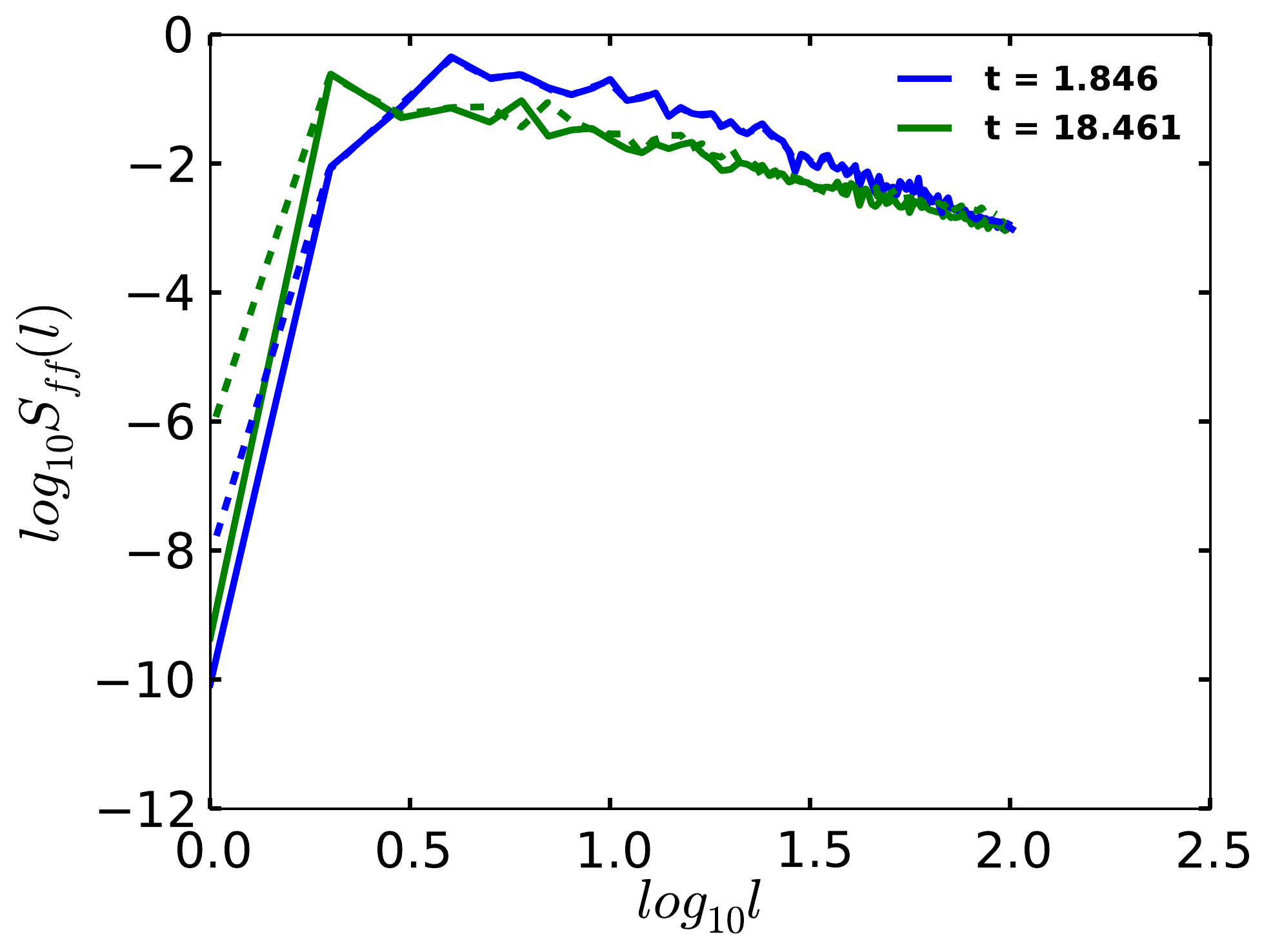}\caption{Vorticity power spectra. The solid line are the DEC results (present study), and the dashed lines are the CLAM results with $n_{g}=1024$  \citep{dritschel2015late}. \label{fig:power_spectra_present_plus_dritschel}}
\end{figure}

\begin{figure}
\centering{}\includegraphics[scale=0.4]{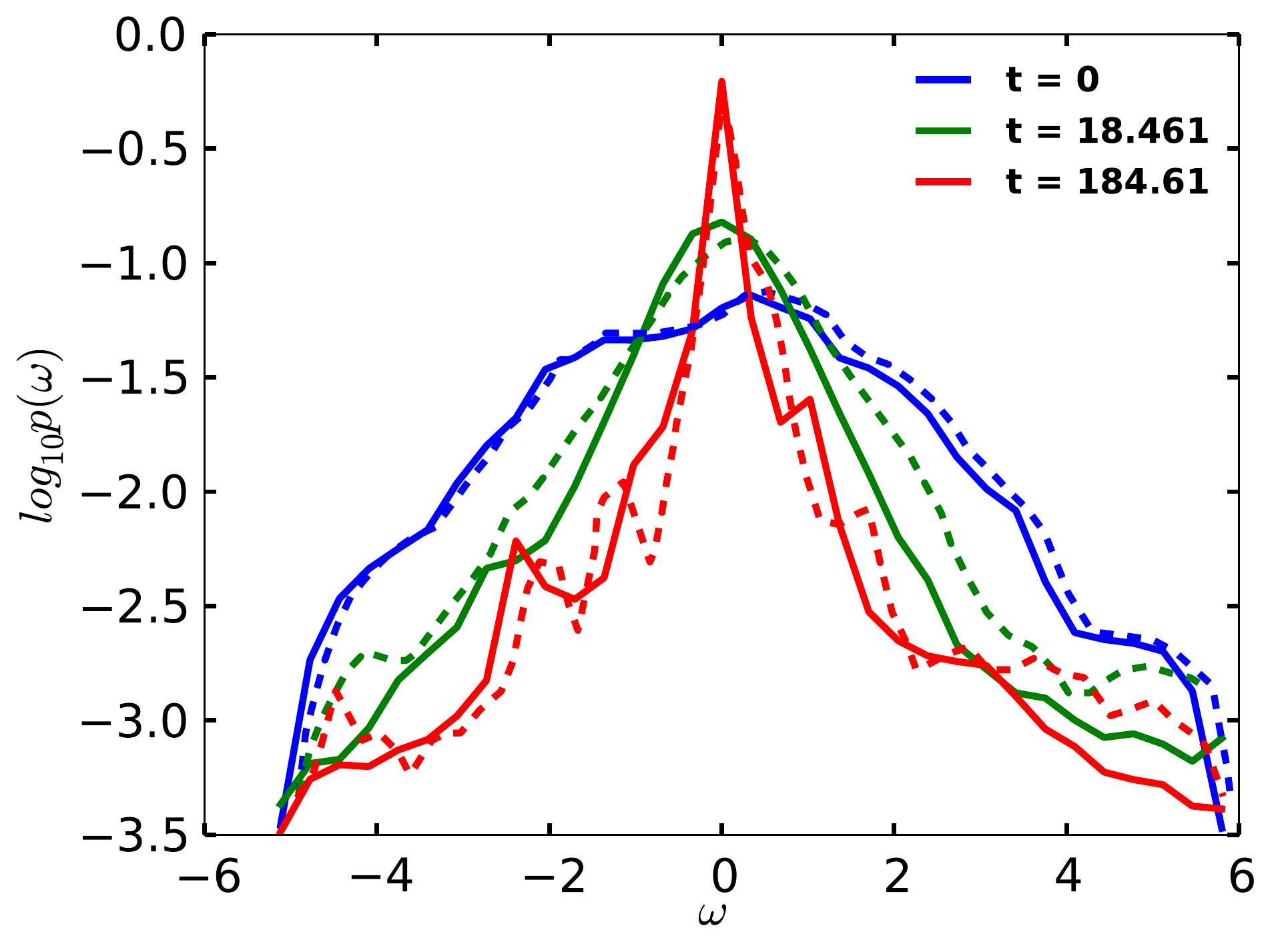}\caption{Vorticity probability density. The solid line are the DEC results (present study), and the dashed lines are the CLAM results with $n_{g}=1024$  \citep{dritschel2015late}. \label{fig:pd_omega_present_plus_dritschel}}
\end{figure}

\begin{figure}
\centering{}\includegraphics[scale=0.4]{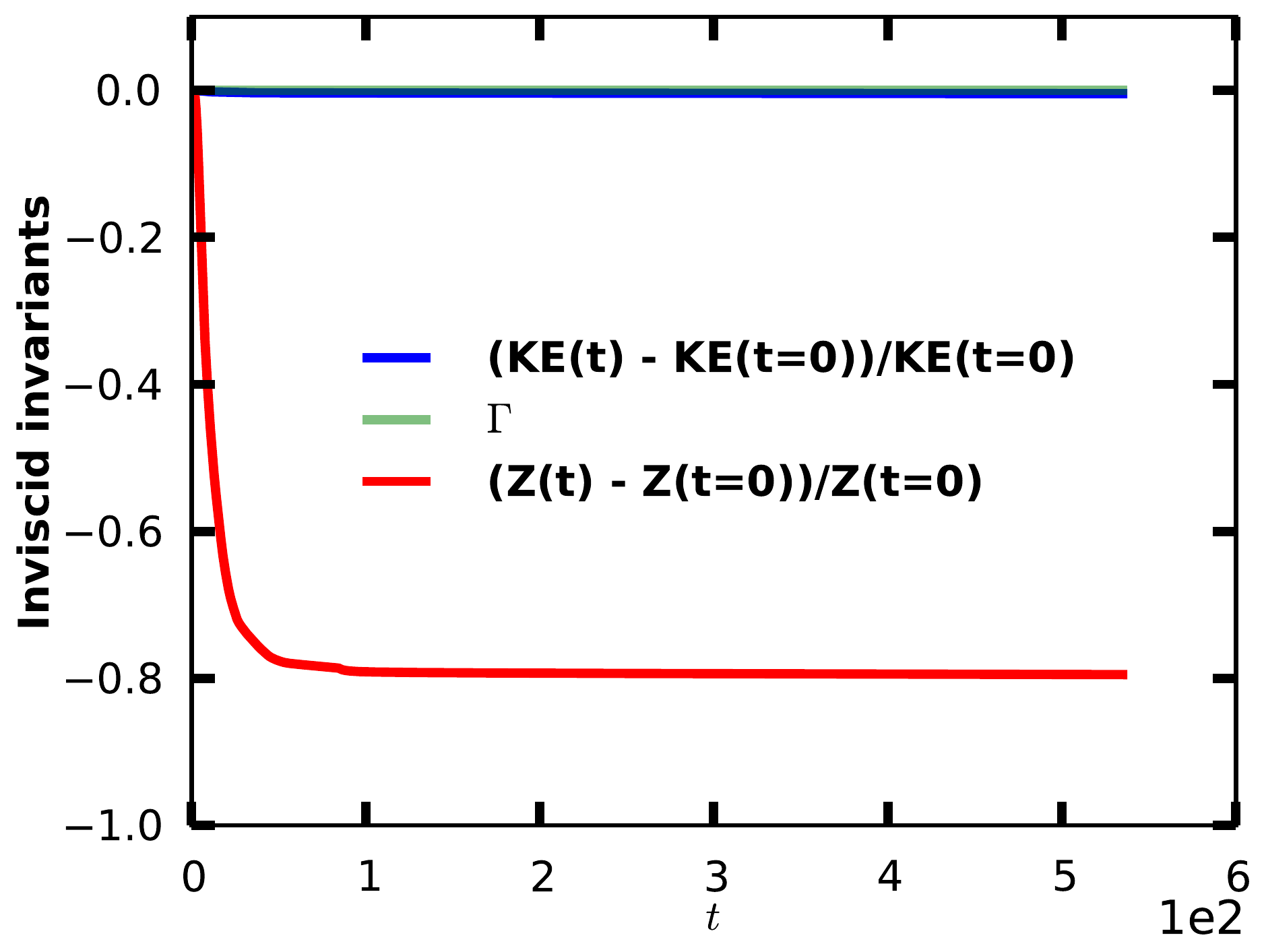}\caption{Inviscid invariants as a function of time. The circulation is preserved to the machine precision, the total kinetic energy and enstrophy decay by 0.5\% and 80\%, respectively, due to the presence of hyperviscosity term in the momentum equation. \label{fig:inviscid_invariants}}
\end{figure}

We investigate the time evolution of vorticity field from the same random initial vorticity field as used in Dritschel {\it et al.}\cite{dritschel2015late} (see figure \ref{fig:ic_present_plus_dritschel}), and compare our results with theirs and that of Modin \& Viviani\cite{modin_viviani_2020}. Our simulations are based on DEC, whereas Dritschel {\it et al.}\cite{dritschel2015late} used  Geodesic Grid Method (GGM), which is approximately second-order accurate, and  Combined Lagrangian Advection Method (CLAM), which is approximately sixteen times finer in each direction than the GGM method for a given mesh,  for their simulations. Modin \& Viviani\cite{modin_viviani_2020} used a casimir preserving scheme. The GGM, CLAM, and DEC methods employed, whereas the casimir preserving scheme did not employ the hyperviscosity in the momentum equation. The vorticity evolution at early times (see figure \ref{fig:t_4_present_plus_dritschel}) is nearly indistinguishable in all simulations (see Modian \& Viviani\cite{modin_viviani_2020} for their results). By $t=18.461$ ($t$ is the eddy turnover time), the vorticity filaments thin exponentially and the vorticity gradients grow exponentially.  The numerical simulations at finite resolution cannot follow this behavior \citep{dritschel2015late}. Although, the vorticity field is not identical, it is qualitatively similar in all simulations (see figure \ref{fig:t_40_present_plus_dritschel}). Similarly, at $t=184.61$ the vorticity field is qualitatively similar in all simulations (see figure \ref{fig:t_400_present_plus_dritschel}).  The vorticity power spectra (see equation (\ref{eq:sffl}) and figure \ref{fig:power_spectra_present_plus_dritschel}), and vorticity probability density (see equation (\ref{eq:pd_vorticity}) and figure \ref{fig:pd_omega_present_plus_dritschel}) are also in agreement with that in Dritschel {\it et al.}\cite{dritschel2015late}. Figure \ref{fig:inviscid_invariants} shows the inviscid invariants as a function of time. The total circulation $\Gamma$ is conserved to the machine precision. We employ a biharmonic viscosity as a means to filter and dissipate very high wavenumber content, and less than 0.5\% of the total kinetic energy dissipates by $t=500$.  About 0.66\% of the total enstrophy $Z$ is lost by $t=1.846$. By $t=18.461$, about 64\% of the total enstrophy dissipates. At $t=184.61$ the enstrophy decays to about 20\% of its initial value. A similar amount of dissipation has also been reported in Dritschel {\it et al.}\cite{dritschel2015late}. Further verification and validation of the DEC method is discussed in the references\onlinecite{mohamed2016discrete,jagad2020primitive,PhysRevFluids.5.044701}.

\section{Segmentation Algorithm for Identifying Individual Vortical
Structures}\label{appB}
The pseudo-code for the algorithm to identify individual vortices are as follows.
\begin{itemize}
\item Read and store the connecting / neighboring nodes to each node in the mesh. 
\item The vorticity magnitude at each node is known from the solution. Chose a suitable threshold value for the vorticity magnitude. Flag the nodes greater than or equal to threshold value as 1 and include them in a list. Flag the remaining nodes as 0.
\item Scroll / loop through the list of nodes with flag = 1:
\begin{itemize}
\item If the flag of the node is 1 include this node in the node list for a vortical structure. Set the nfound count to 1. 
\item While nfound is greater than 0: 
\begin{itemize}
\item Call the function Fill(node list). 
\item Update the node list from the list of connecting / neighboring nodes received from the function Fill. Also, updated nfound corresponding to the list of connecting / neighboring nodes is received from the function Fill.
\end{itemize}
\end{itemize}
\end{itemize}

Fill function 
\begin{itemize}
\item Scroll through the nodes in the node list (supplied via a function argument):
\begin{itemize}
\item Set the flag of the node to 0.
\item Set the nfound count to 0.
\item Scroll through the connecting / neighboring nodes to the node being scrolled in the outer loop. If the flag of the neighboring node is 1, include this node in the list of connecting nodes and increase the nfound count by 1, flag this node as 0. 
\item Return the list of connecting nodes and nfound to the calling function. 
\end{itemize}
\end{itemize} 

%\section*{References}
%\nocite{*}
\bibliography{paper}

\end{document}